\documentclass[aps,amssymb,floatfix,superscriptaddress,preprintnumbers,nofootinbib]{revtex4}

\usepackage{graphicx}
\usepackage{bm}
\usepackage{amsmath}
\usepackage{amssymb}
\usepackage{amsfonts}
\usepackage{float}
\usepackage{hyperref}
\usepackage{dsfont}  
\usepackage{slashed}  
\usepackage{booktabs}
\usepackage{multirow}
\usepackage{subfigure}
\usepackage[sort&compress]{natbib}
\usepackage{xcolor}
\usepackage{ulem}

\newcommand{\be}{\begin{equation}}  
\newcommand{\ee}{\end{equation}}

\newcommand{\bea}{\begin{eqnarray}}
\newcommand{\eea}{\end{eqnarray}}

\newcommand{\Slash}[1]{{\ooalign{\hfil/\hfil\crcr$#1$}}}

\newcommand{\nn}{\nonumber\\}
\newcommand{\beq}{\begin{eqnarray}}
\newcommand{\eeq}{\end{eqnarray}}

\newcommand{\UV}{\text{\tiny{UV}}}
\newcommand{\IR}{\text{\tiny{IR}}}

\begin{document}

\title{Chiral and trace anomalies in Deeply Virtual Compton Scattering II: \\ 
QCD factorization and beyond}

\author{Shohini Bhattacharya}
\email{sbhattach@bnl.gov}
\affiliation{RIKEN BNL Research Center, Brookhaven National Laboratory, Upton, NY 11973, USA}

\author{Yoshitaka Hatta}
\email{yhatta@bnl.gov}
\affiliation{Physics Department, Brookhaven National Laboratory, Upton, NY 11973, USA}
\affiliation{RIKEN BNL Research Center, Brookhaven National Laboratory, Upton, NY 11973, USA}

\author{Werner Vogelsang}
\email{werner.vogelsang@uni-tuebingen.de}
\affiliation{Institute for Theoretical Physics, T\"{u}bingen University, Auf der Morgenstelle 14, 72076 T\"{u}bingen, Germany}

\begin{abstract}

\vspace*{0.4cm}
\noindent 
We extend the discussion of the recently discovered `anomaly poles' in QCD Compton scattering.  
We perform the complete one-loop calculation of the Compton amplitude using momentum transfer $t$ as the regulator of collinear divergences. In the gluon channel, we confirm the presence of poles $1/t$  in both the real and imaginary parts of the amplitude. In the quark channel, we find unexpected infrared single $1/\epsilon$ and double $1/\epsilon^2$ poles. We then perform the one-loop calculation of the leading-twist quark generalized parton distributions (GPDs) for quark and gluon external states with the same regulators and 
find that all these singular terms can be systematically absorbed into the GPDs, showing that QCD factorization is restored to this order. Having established this, we discuss the fate of the $1/t$ poles. We argue that they become the  nonperturbative building blocks of GPDs that encode the chiral and trace anomalies of QCD, in a way consistent with the known constraints these  anomalies impose on the nucleon axial and gravitational form factors. The scope of research on GPDs can therefore be expanded to address the manifestation and implications of quantum anomalies in high-energy exclusive processes.

\end{abstract}

\maketitle

\section{Introduction}
The past several years have witnessed significant progress in the higher-order calculation of Deeply Virtual Compton Scattering (DVCS). In the flavor-nonsinglet channel, the three-loop evolution equation for the generalized parton distributions (GPDs) has been derived \cite{Braun:2017cih} together with the  two-loop coefficient functions \cite{Braun:2020yib}. In the flavor-singlet channel, the two-loop coefficient functions for DVCS have been recently calculated  \cite{Braun:2022bpn} and even higher order resummation effects have been studied    
\cite{Schoenleber:2022myb}. These developments are on a steady path toward achieving the NNLO accuracy in DVCS  that is required for precision GPD studies at the future Electron-Ion Collider (EIC) \cite{AbdulKhalek:2021gbh}. 

Meanwhile, in a previous paper \cite{Bhattacharya:2022xxw}, we have explored  a new approach to compute the NLO corrections in DVCS, following an earlier suggestion in polarized deep inelastic scattering (DIS)  \cite{Tarasov:2020cwl,Tarasov:2021yll}. 
The key ingredient is to use momentum transfer $t=(P_1-P_2)^2$ as an infrared cutoff to regulate  the collinear divergence,  instead of the usual dimensional regularization. Previously in the calculation of the Compton amplitude in the Bjorken limit, the variable $t$ had always been neglected when computing partonic amplitudes.  Naively, one would expect  that the only new effect of introducing nonzero $t$ would be to generate  higher twist corrections of order $|t|/Q^2\ll 1$ where $Q^2$ is the photon virtuality. However,  our explicit calculations with nonzero $t$  have  revealed  `anomaly poles' $1/t$ which had not been detected in the previous calculations performed at $t=0$ \cite{Ji:1997nk,Belitsky:1997rh,Mankiewicz:1997bk,Pire:2011st,Bertone:2022frx}, but are consistent with the result in \cite{Tarasov:2020cwl}. Moreover, these poles are accompanied by certain twist-{\it four} GPDs but without an expected suppression factor $1/Q^2$. (Rather, $1/Q^2$ has been replaced by $1/t$.) In fact, they can be interpreted as the manifestations of the QCD chiral \cite{Jaffe:1989xy,Tarasov:2020cwl,Tarasov:2021yll} and trace \cite{Bhattacharya:2022xxw} anomalies in high energy scattering. The finding thus points towards a novel connection between the study of GPDs and phenomena associated with quantum anomalies such as chiral symmetry breaking and confinement.

At face value, the emergence of poles is in apparent contradiction with  the QCD factorization theorem \cite{Collins:1998be,Ji:1997nk} which states that  the QCD Compton scattering amplitude factorizes into the perturbatively calculable coefficient functions and the nonperturbative twist-two GPDs up to higher-twist corrections of order $1/Q^2$. However, we have already suggested in  \cite{Bhattacharya:2022xxw} that the poles found in the one-loop calculation may be absorbed into  the twist-two GPDs as a part of the infrared subtraction procedure.  The purpose of this paper is to fully demonstrate that this is indeed the case. We first perform a complete calculation of the Compton amplitude with nonzero $t$ at 
one loop, both in the quark and gluon channels, in 
both the polarized and unpolarized sectors, and for the real and imaginary parts of the amplitude. (In \cite{Bhattacharya:2022xxw}, we only calculated the imaginary part in the gluon channel.) In the gluon channel, we find $1/t$ poles also in the real part. 
Surprisingly, in the quark channel, we find uncancelled infrared single $1/\epsilon$ and double $1/\epsilon^2$ poles.  
We next perform the corresponding one-loop calculation of the unpolarized and polarized quark GPDs for free quark and gluon external states at finite $t$ and show that all the singular terms  can be systematically absorbed.  

Therefore, at least to one loop, the emergence of $1/t$ poles does not contradict the QCD factorization theorem. The calculation with nonzero $t$ may be regarded as an alternative factorization scheme. 
Having established this, we shift our focus to  the fate of the $1/t$ poles absorbed into the twist-two GPDs.  It is well known that the chiral and trace anomalies impose constraints on the nucleon axial and gravitational form factors, respectively. Since these form factors are certain moments of the twist-two GPDs, there must be corresponding constraints  directly on GPDs \cite{Bhattacharya:2022xxw}. 
A preliminary discussion of this has been already presented in \cite{Bhattacharya:2022xxw}. Our extended treatment here will lend more support to the idea that this new scheme can uniquely address such profound aspects of QCD in GPD studies.

\section{Compton scattering}
The amplitude for QCD Compton scattering off a proton target,  $\gamma^*(q_1)p(P_1)\to \gamma^*(q_2)p(P_2)$, is given by  
\beq
T^{\mu\nu} &=&i\int d^4y e^{iq\cdot y}\langle P_2|{\rm T}\{j^\mu(y/2)j^{\nu}(-y/2)\}|P_1\rangle,
\eeq
where $j^\mu=\sum_q e_q \bar{\psi}_q\gamma^\mu \psi_q$ is the electromagnetic current and $q^\mu=\frac{q^\mu_1+q^\mu_2}{2}$ is the average of the incoming and outgoing photon momenta. The momentum transfer is denoted as $t=l^2$ where $l^\mu=P_2^\mu-P_1^\mu=q_1^\mu-q_2^\mu$. We  introduce 
the generalized Bjorken variable $x_B$ and the skewness parameter $\xi$,
\beq
 x_B=\frac{Q^2}{2P\cdot q}, \quad \xi=\frac{q_2^2-q_1^2}{2P\cdot q} \approx -\frac{l^+}{2P^+} , \label{xb}
\eeq
where $Q^2=-q^2$ is the photon virtuality and $P^\mu=\frac{P_1^\mu+P_2^\mu}{2}$. In DVCS, $q_2^2=0$ and  $x_B\approx \xi$, but we shall keep general $x_B$ and $\xi$ throughout the paper.

In the generalized Bjorken limit $Q^2,\, 2P\cdot q\to \infty$ with $x_B$,  $t$ fixed and $Q^2\gg |t|$,  the Compton amplitude can be expanded as \cite{Diehl:2003ny,Belitsky:2005qn}
\beq
T^{\mu\nu} = \frac{g^{\mu\nu}_\perp}{2P^+} \bar{u}(P_2)\left[\gamma^+ {\cal H}+ \frac{i\sigma^{+\nu}l_\nu}{2M}{\cal E}\right]u(P_2) -i\frac{\epsilon^{\mu\nu}_\perp}{2P^+}\bar{u}(P_2)\left[\gamma^+\gamma_5 \tilde{{\cal H}}+ \frac{\gamma^5 l^+}{2M}\tilde{{\cal E}}\right]u(P_2) +\cdots \, , \label{tmunu}
\eeq
where $M$ is the proton mass. $g_\perp^{\mu\nu}$ and $\epsilon_\perp^{\mu\nu}$ are transverse projectors such that $g^{ij}_\perp = -\delta^{ij}$ and  $\epsilon^{ij}_\perp=\epsilon^{ij}$  for transverse indices $i,j=1,2$ and the other components are zero. Our convention is  $\gamma_5=i\gamma^0\gamma^1\gamma^2\gamma^3$ and $\epsilon^{0123}=\epsilon^{-+12}=\epsilon^{12}=1$. 
The ellipsis in (\ref{tmunu}) stand for the contributions from the (generalized) longitudinal  structure function and the so-called gluon transversity GPD. As observed in \cite{Bhattacharya:2022xxw}, they are not sensitive to anomalies and are thus left for future work. 

According to QCD factorization, the Compton form factors ${\cal H}$, ${\cal E}$, $\tilde{\cal H}$  and $\tilde{\cal E}$ can be written as convolutions of nonperturbative GPDs and partonic hard-scattering amplitudes. The latter are commonly calculated in dimensional regularization in $d=4-2\epsilon$ dimensions, with $\epsilon$ regularizing both  ultraviolet (UV) and infrared (IR) divergences. We shall also work in $d$ dimensions, since individual diagrams contain UV divergences. But we regularize the collinear singularity by introducing the physical variable $t$ at the partonic level.  Such a calculation is safely justified when $\sqrt{|t|}\gg \Lambda_{\rm QCD}$ (still keeping $Q \gg \sqrt{|t|}$),\footnote{ We consider large but finite $\sqrt{|t|}$, so we are still in the generalized Bjorken limit $|t|\ll Q^2\to \infty$ and the usual factorization in terms of the twist-two GPDs is expected. This is different from so-called wide angle Compton scattering   (see, e.g., \cite{Radyushkin:1998rt}) where $Q=0$ and $s\sim -t \sim -u$. In our calculation, $s\sim Q^2$ and we systematically neglect terms of order $t/Q^2\sim t/s$. }  but we shall eventually be interested in the region $\sqrt{|t|}\sim \Lambda_{\rm QCD}$.  
The result of the one-loop calculation  can be summarized in the form   
\beq
\begin{pmatrix} {\cal H}(x_B,\xi,t) \\ {\cal E}(x_B,\xi,t)\end{pmatrix} &=& \sum_q e_q^2 \int_{0}^1 dx \Biggl[ \left(C_0(x,x_B) + \frac{\alpha_s}{2\pi}C^q_1(x,x_B,\xi)\right)\begin{pmatrix} H_q(x,\xi,t)-H_q(-x,\xi,t) \\ E_q(x,\xi,t)-E_q(-x,\xi,t) \end{pmatrix}  \nn 
&& \qquad \qquad \qquad +  \frac{\alpha_s}{2\pi}  C^g_1(x,x_B,\xi)\begin{pmatrix} H_g(x,\xi,t) \\ E_g(x,\xi,t) \end{pmatrix}  \nn  && \qquad  \qquad \qquad + \frac{\alpha_s}{2\pi}\frac{M^2}{t} A(x,x_B,\xi) \begin{pmatrix} {\cal F}(x,\xi,t) \\ -{\cal F}(x,\xi, t)\end{pmatrix} \Biggr] + {\cal O}(1/Q^2)+{\cal O}(\alpha_s^2), \label{sym}
\eeq
\beq
\begin{pmatrix} \tilde{{\cal H}}(x_B,\xi,t) \\ \tilde{{\cal E}}(x_B,\xi,t)\end{pmatrix} &=& \sum_q e_q^2\int_{0}^1 dx\Biggl[\left(\tilde{C}_0(x,x_B)+\frac{\alpha_s}{2\pi}\tilde{C}^q_1(x,x_B,\xi)\right)\begin{pmatrix} \tilde{H}_q(x,\xi,t)+\tilde{H}_q(-x,\xi,t) \\ \tilde{E}_q(x,\xi,t)+\tilde{E}_q(-x,\xi,t) \end{pmatrix}  \nn 
&& \qquad \qquad \qquad + \frac{\alpha_s}{2\pi}\tilde{C}_1^g(x,x_B,\xi)\begin{pmatrix} \tilde{H}_g(x,\xi,t) \\ \tilde{E}_g(x,\xi,t) \end{pmatrix}  \nn  && \qquad \qquad \qquad +   \frac{\alpha_s}{2\pi}\frac{M^2}{t} \tilde{A}(x,x_B,\xi) \begin{pmatrix} 0 \\ \tilde{\cal F}(x,\xi, t)\end{pmatrix} \Biggr]+{\cal O}(1/Q^2)+{\cal O}(\alpha_s^2), \label{asym}
\eeq
where the notations for the twist-two quark and gluon GPDs $H_{q,g}$, $E_{q,g}$, $\tilde{H}_{q,g}$, $\tilde{E}_{q,g}$ are standard \cite{Diehl:2003ny,Belitsky:2005qn}. 
(The gluon GPDs are normalized as $H_g(x)=xG(x)$ and $\tilde{H}_g(x)=x\Delta G(x)$ in the forward limit, where $G$, $\Delta G$ are the unpolarized and polarized gluon PDFs.) Note that (\ref{sym}) and (\ref{asym})  are still in their `unsubtracted' forms in the sense that some of the coefficients contain divergences  in the formal limit $t\to 0$. Their subtraction is rather unconventional, and to elaborate on this is one of the main objectives of our paper.

The leading-order coefficient functions are well known:
\begin{align}
C_0\left(x,x_B\right) &= \frac{1}{x-x_B+i\epsilon}+\frac{1}{x+x_B-i\epsilon}, \nn 
\tilde{C}_0\left(x,x_B\right) &=\frac{1}{x-x_B+i\epsilon}-\frac{1}{x+x_B-i\epsilon}. \label{leading}
\end{align}
The one-loop corrections, $C_1^q$ etc., have the following generic structure:
\begin{align}
C^q_1(x,x_B,\xi)&=\frac{C_F}{x} \left(\kappa_{qq}(\hat{x},\hat{\xi}) \ln \frac{Q^2}{-l^2} +\delta C^q_1(\hat{x},\hat{\xi})\right), \qquad \tilde{C}^q_1(x,x_B,\xi)= \frac{C_F}{x}\left(\tilde{\kappa}_{qq}(\hat{x},\hat{\xi}) \ln \frac{Q^2}{-l^2} +\delta \tilde{C}^q_1(\hat{x},\hat{\xi})\right), \label{log1}
\\[0.3cm]
C^g_1(x,x_B,\xi)&= \frac{2T_R}{x^2}\left(\kappa_{qg}(\hat{x},\hat{\xi}) \ln \frac{Q^2}{-l^2} +\delta C^g_1(\hat{x},\hat{\xi})\right), \qquad \tilde{C}^g_1(x,x_B,\xi)=\frac{2T_R}{x^2}\left( \tilde{\kappa}_{qg}(\hat{x},\hat{\xi}) \ln \frac{Q^2}{-l^2} +\delta \tilde{C}^g_1(\hat{x},\hat{\xi})\right), \label{log2}
\end{align}
where $C_F=\frac{4}{3}$ and $T_R=\frac{1}{2}$ are the usual color factors. We have introduced the partonic variables $\hat{x}=\frac{x_B}{x}$ and $\hat{\xi}=\frac{\xi}{x}$, and  set the $\overline{\rm MS}$ renormalization scale to be $4\pi e^{-\gamma_E}\mu^2=Q^2$. 
The logarithm $\ln \frac{Q^2}{-l^2}$ originates from the collinear singularity and replaces the $-\tfrac{1}{\epsilon_{\IR}}$ pole  in the usual calculation in dimensional regularization with $t=0$. The coefficients $\kappa$, $\tilde{\kappa}$ are fixed by the evolution equation of GPDs and  must agree with the known results in the literature.   On the other hand, the coefficient functions $\delta C_1^q$, $\delta C_1^g$, $\delta \tilde{C}_1^q$ and $\delta\tilde{C}_1^g$ are potentially scheme dependent.  The results in the $\overline{\rm MS}$ scheme  can be found in \cite{Ji:1997nk,Belitsky:1997rh,Mankiewicz:1997bk,Belitsky:2005qn}. 
Note that, somewhat at variance with the previous
literature, we have used the reflection symmetry in  $x$ to restrict the $x$-integral to the region $0<x<1$. Namely,  $\tilde{C}_0$, $\tilde{C}_1^q$, $C_1^g$, $H_g$, $E_g$ are even functions and  $C_0$, $C_1^q$, $\tilde{C}_1^g$  $\tilde{H}_g$, $\tilde{E}_g$ are odd functions, respectively, under $x\to -x$. This is convenient for the discussion below.

Eqs.~(\ref{sym}) and (\ref{asym}) resemble the usual structure dictated by the QCD factorization theorem  
except for the `anomaly pole' terms  proportional to $1/t$. These poles are accompanied by the twist-four gluon  GPDs
${\cal F}$ and $\tilde{\cal F}$  defined as \cite{Hatta:2020iin,Tarasov:2020cwl,Hatta:2020ltd,Radyushkin:2021fel,Radyushkin:2022qvt} 
\begin{align}
{\cal F} (x,\xi,t)&\equiv \frac{-4xP^+}{M} \int \frac{dz^-}{2\pi} e^{ixP^+z^-} \frac{ \langle P_2|F^{\mu\nu}(-z^-/2)WF_{\mu\nu}(z^-/2)|P_1\rangle}{\bar{u}(P_2)u(P_1)} \, , \label{four}
\\[0.3cm]
\tilde{{\cal F}}(x,\xi,t)&\equiv  \frac{iP^+}{M}\int \frac{dz^-}{2\pi} e^{ixP^+z^-} \frac{\langle P_2|F^{\mu\nu}(-z^-/2)W\tilde{F}_{\mu\nu}(z^-/2)|P_1\rangle}{\bar{u}(P_2)\gamma_5u(P_1)} \, , \label{two}
\end{align}
where $W$ is the straight Wilson line between $[-z^-/2,z^-/2]$. We have changed the normalization with respect to \cite{Bhattacharya:2022xxw} in order to make these distributions dimensionless. Despite involving twist-four GPDs, these terms are not suppressed by $1/Q^2$ and apparently cause problems in the forward limit $t\to 0$. As discussed in \cite{Bhattacharya:2022xxw} and will be further elaborated later, the emergence of poles and their fate may shed new light on the nonperturbative structure of GPDs, connecting to deep issues such as  chiral symmetry breaking and the origin of hadron masses.

\section{Calculations}
\begin{figure}[t]
\centering
\includegraphics[width = 15cm]{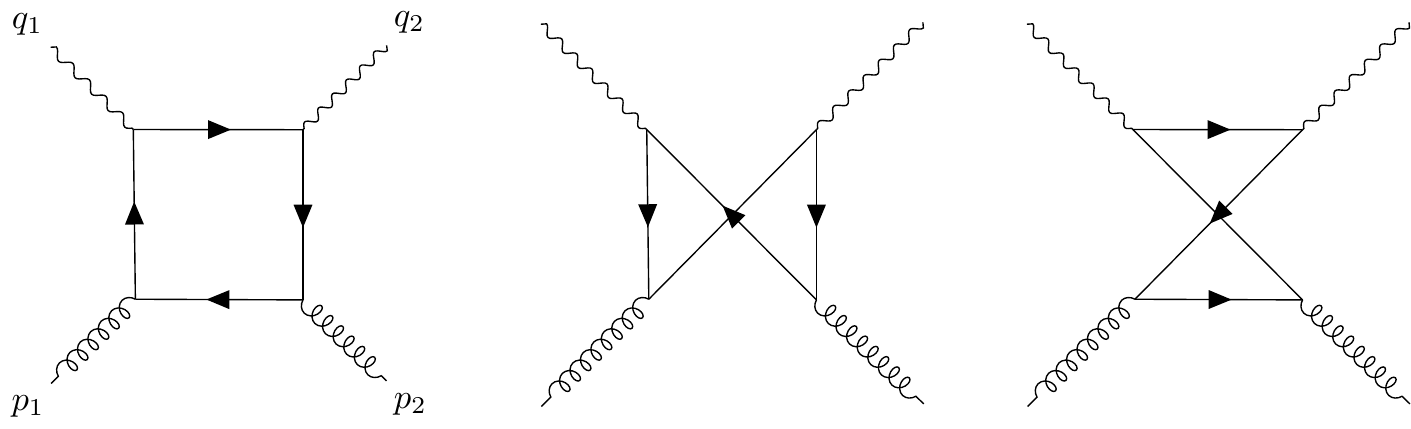}
\caption{Diagrams for the subprocess $\gamma^* g\to \gamma^*g$ in Compton scattering.}
\label{fig1}
\end{figure}
\begin{figure}[t]
\centering
\includegraphics[width = 18.5cm]{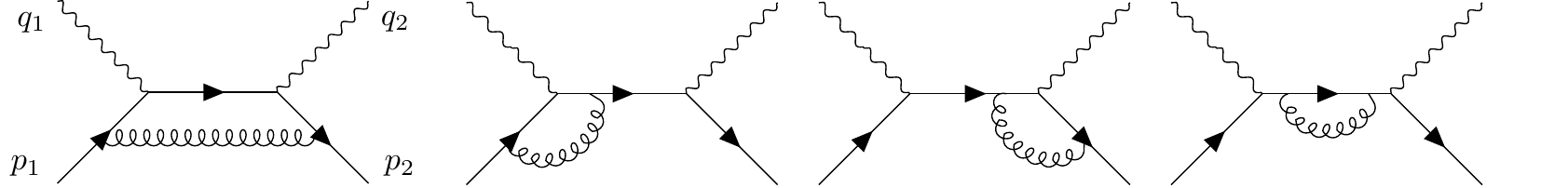}
\caption{Diagrams for the subprocess $\gamma^* q \to \gamma^* q$ in Compton scattering. Diagrams with photon lines crossed are not shown. }
\label{fig2}
\end{figure}
In this section, we outline our calculation of the perturbative corrections to Compton scattering at one-loop order. The relevant Feynman diagrams for the subprocess initiated by the gluons are shown in Fig.~\ref{fig1} and the ones initiated by the quarks are shown in Fig.~\ref{fig2}. For the latter case, we choose to work in Feynman gauge. (We have also worked in a general covariant gauge and confirmed that the final results are independent of the gauge as it should be.) As in Ref.~\cite{Bhattacharya:2022xxw}, we parameterize the incoming and outgoing momenta as 
\beq
q_1=q+\frac{l}{2}, \quad q_2=q-\frac{l}{2}, \quad p_1=p-\frac{l}{2}, \quad p_2=p+\frac{l}{2}.
\eeq
We also define the partonic versions of the Bjorken and skewness variables (\ref{xb}) as
\beq
\hat{x}= \frac{Q^2}{2p\cdot q} = \frac{x_B}{x}, 
\qquad \hat{\xi} 
= \frac{\xi}{x} 
= -\hat{x} \frac{ q\cdot l}{Q^2} .
\eeq
The incoming and outgoing partons are assumed to be massless, $p_1^2=p_2^2=0$, which leads to the conditions $p^2=-l^2/4$ and $p\cdot l=0$. The virtuality of the photons can be written as
\beq
q_1^2= -Q^2\frac{\hat{x}+\hat{\xi}}{\hat{x}}+\frac{l^2}{4} , \qquad q_2^2= -Q^2\frac{\hat{x}-\hat{\xi}}{\hat{x}}+\frac{l^2}{4}. \label{timelike}
\eeq 
We will abbreviate the polarization vectors for the gluons in Fig.~\ref{fig1} as $\epsilon^\mu(p_1)\equiv \epsilon_1^\mu$ and $\epsilon^{*\mu}(p_2)\equiv \epsilon_2^{*\mu}$.

The collinear singularity in the above diagrams will be regularized by $t=l^2$. We emphasize that, in the present `handbag' approximation, $t=(p_2-p_1)^2=(P_2-P_1)^2$ is the same at the hadronic and partonic levels. However, we still have to work in $d=4-2\epsilon$ dimensions because the individual diagrams will contain UV divergences in the real part. At the same time, working in $d$-dimensions also helps to check if there are any leftover IR divergences that are not regularized by nonzero $t$ alone. This point will be relevant for the quark-channel diagrams in Fig.~\ref{fig2}. Our convention is that, if $\epsilon$ is used for the UV divergences, then $\epsilon \rightarrow \epsilon_{\UV} > 0$, while if it is used for the IR divergences then $\epsilon \rightarrow \epsilon_{\IR} < 0$.

In the following, we shall refer to the two terms in (\ref{tmunu}) as the symmetric and antisymmetric parts of the Compton amplitude, respectively. 
The symmetric part can be extracted with the help of  the projector
\begin{align}
g^{\mu\nu}_\perp & =g^{\mu\nu}+\frac{1}{q^2(1+\gamma^2)}\left(q^\mu-\frac{q^2}{p\cdot q}p^\mu\right)\left(q^\nu - \frac{q^2}{p\cdot q}p^\nu\right) -\frac{q^\mu q^\nu}{q^2} \, , \qquad \gamma^2=-\frac{p^2q^2}{(p\cdot q)^2}=\frac{l^2q^2}{4(p\cdot q)^2} ,
\label{e:sym_proj}
\end{align}
such that
\beq
g^\mu_{\perp\mu}=d-2=2(1-\epsilon), \qquad {\cal H},{\cal E}\sim \frac{1}{2(1-\epsilon)}g_{\mu\nu}^\perp T^{\mu\nu} \, .
\eeq
For the antisymmetric part, we use the projector $\epsilon^{\alpha p \mu \nu} \equiv \epsilon^{\alpha\beta \mu\nu}p_\beta$.

We evaluate the above diagrams with the help of the Mathematica package `Package-X'~\cite{Patel:2015tea}.
Below we first discuss the main features of our results specific to the gluon and quark channels. The complete results will then be presented in Sec.~\ref{s:results}.

\subsection{Gluon channel}
Our results feature (i) a ${1}/{l^2}$ pole and (ii) a logarithm $\ln({Q^{2}/}{-l^2})$, both arising from the first and third diagrams in Fig.~\ref{fig1}. For the symmetric case, the UV poles from the first and third diagrams add up to cancel the one arising from the second diagram. For the antisymmetric case, the UV poles from the first and third diagram cancel. 
There are no leftover ${1}/{\epsilon_{\IR}}$ divergences, meaning that $l^2 \neq 0$ functions as a genuine  regulator of collinear divergences.

In the symmetric case, the result for the one-loop Compton scattering amplitude  with external gluon polarization vectors $\epsilon_1$, $\epsilon_2^{*}$ (Fig.~\ref{fig1}) has the following generic structure:
\beq
-\epsilon_1\cdot \epsilon^*_2 \left(A\ln \frac{Q^2}{-l^2}+B\right) +\frac{C}{l^2}\epsilon_1 \cdot l \epsilon_2^*\cdot l  =
-\epsilon_1\cdot \epsilon^*_2 \left(A\ln \frac{Q^2}{-l^2}+B-\frac{C}{2}\right) +\frac{C}{l^2}\left(\epsilon_1 \cdot l \epsilon_2^*\cdot l-\frac{\epsilon_1\cdot \epsilon_2^*}{2}l^2\right)  ,
\label{symg}
\eeq
where $A,B,C$ are coefficients that depend on $\hat{x}$ and $\hat{\xi}$.
In the asymmetric case, we find instead 
\beq
i\epsilon^{\alpha p \epsilon^*_2 \epsilon_1}\left(\tilde{A}\ln \frac{Q^2}{-l^2}+\tilde{B}\right) +\tilde{C}\frac{l^\alpha}{l^2} i\epsilon^{\epsilon_1\epsilon_2^* l p}, \label{asymg}
\eeq
where $\epsilon^{\epsilon_1\epsilon_2^* l p}\equiv \epsilon^{\mu\nu\rho\lambda} \epsilon_{1\mu}\epsilon^*_{2\nu} l_\rho p_\lambda$. 
The first terms in (\ref{symg}) and (\ref{asymg}) can be interpreted as the usual one-loop corrections to the Compton amplitude $\sim C_1^g H_g, \tilde{C}_1^g\tilde{H}_g$ through the identifications 
\beq
 -\epsilon_1\cdot \epsilon_2^*\sim \frac{\langle p_2|F^{+\mu}F^{\ +}_{\mu}|p_1\rangle}{1-\hat{\xi}^2}, \qquad  
 i \epsilon^{+p \epsilon^*_2 \epsilon_1}\sim \frac{\langle p_2| i F^{+\mu}\tilde{F}^{\ +}_{ \mu}|p_1\rangle}{1-\hat{\xi}^2}. \label{matr}
 \eeq
 However, the second terms  in (\ref{symg}) and (\ref{asymg})  cannot be attributed to twist-two GPDs. Their structures can  only arise from the twist-four operators $F^{\mu\nu}F_{\mu\nu}$ and $F^{\mu\nu}\tilde{F}_{\mu\nu}$
  \beq
\epsilon_1 \cdot l \epsilon_2^*\cdot l - \frac{\epsilon_1\cdot \epsilon^*_2}{2}l^2\sim \langle p_2|F^{\mu\nu}F_{\mu\nu}|p_1\rangle,
\qquad 2i\epsilon^{\epsilon_1\epsilon_2^* l p}\sim \langle p_2|iF^{\mu\nu}\tilde{F}_{\mu\nu}|p_1\rangle,
\label{ele}
\eeq
and this is how the twist-four GPDs  (\ref{four}), (\ref{two}) come into play. 
It should be mentioned, however, that the present argument only concerns  the two-parton matrix element of the operators $FF$ and $F\tilde{F}$. Further justifications from other approaches are desirable. 

\subsection{Quark channel}
In this case, our results do not contain any terms ${1}/{l^2}$. This is consistent with the expectation
that the anomalies, being of purely gluonic nature, should not affect the quark sector, at least at this order.
Quite unexpectedly though, we find (i) double IR poles ${1}/{\epsilon^2_\IR}$, (ii) single IR poles ${1}/{\epsilon_\IR}$, apart from (iii) a logarithm  $\ln({Q^{2}}/{-l^2})$. Besides, we also find (iv) UV poles ${1}/{\epsilon_\UV}$. It is interesting to discuss the origin of these poles.
The UV poles arise from all the diagrams in Fig.~\ref{fig2} excluding the first. To cancel them, we need to include the square root of the self-energy corrections $0=\frac{1}{\epsilon_\UV}-\frac{1}{\epsilon_\IR}$ on the incoming and outgoing (massless) quark lines. This converts the UV poles into single IR poles that add to the 
ones in (ii).  
The double IR poles arise from the first diagram while the single IR poles arise from all diagrams except the fourth. It is interesting to note that, in inclusive DIS, the second and third diagrams also give rise to double poles, canceling the one from the first diagram, but in the present case they do not because the quark lines (after re-absorption of gluons) are massive.
Another interesting feature is that, in the usual $\overline{\rm MS}$ scheme, one obtains the evolution kernel of GPDs as the coefficient of single IR poles for which all the aforementioned diagrams contribute. In our case, we reproduce the kernel as the coefficient of the logarithm  (iii) for which only the first diagram contributes.

In the antisymmetric case, we project the result onto $\bar{u}(p_2)\gamma^+\gamma_5 u(p_1)$ in order to extract the twist-two component. This makes it necessary to specify the treatment of the Dirac matrix in $d\neq 4$ dimensions.
We have used both the `naive' fully anticommuting $\gamma_5$ and the HVBM scheme \cite{tHooft:1972tcz,Breitenlohner:1977hr}. 
For the latter we have computed the Dirac traces using the Mathematica package `Tracer' \cite{Jamin:1991dp}.
The HVBM scheme provides the preferred scheme \cite{Weber:1991wd,Vogelsang:1996im} because, unlike the one with the 
fully anticommuting $\gamma_5$, it is known to be algebraically consistent. Remarkably, however, in the present case 
the result is the same for both schemes. The reason is that all $1/\epsilon$ pole terms we find enter with a part of
the Dirac trace that manifestly gives the same answer for both treatments of $\gamma_5$. 
All further collinear singularities are regularized by the logarithm $\ln({Q^{2}}/{-l^2})$, so that this part of the calculation can essentially be carried out in four dimensions where of course both schemes coincide. The same is true
for the gluonic coefficient function which has no poles in $1/\epsilon$ and can be obtained in four dimensions. 
Hence there are no ambiguities related to the Levi-Civita tensor being an entirely four-dimensional object. 
The issue of the scheme nevertheless will show up later when we discuss the one-loop calculation of the GPDs.

\section{Results}
\label{s:results}

\subsection{One-loop evolution kernels}

The coefficients of the logarithm $\ln({Q^{2}/}{-l^2})$ in (\ref{log1}), (\ref{log2}) are dictated by the evolution of the twist-two GPDs and therefore must agree with the known results  in the literature 
\cite{Ji:1997nk,Belitsky:1997rh,Mankiewicz:1997bk}.   We find that this is indeed the case, meaning that the physical parameter $t$ does the job of regularizing the collinear singularity associated with GPDs.  For completeness, here we  reproduce the results: 
\beq
\kappa_{qq}(\hat{x},\hat{\xi}) 
&=&\frac{3}{2(1-\hat{x})} +\frac{\hat{x}^2+1-2\hat{\xi}^2}{(1-\hat{\xi}^2)(1-\hat{x})} \ln \frac{\hat{x}-1}{\hat{x}}   + \frac{(\hat{x}-\hat{\xi})(1-\hat{x}^2-2\hat{x} \hat{\xi} -2\hat{\xi}^2)}{(1-\hat{x}^2)\hat{\xi} (1-\hat{\xi}^2)} \ln \frac{\hat{x}-\hat{\xi}}{\hat{x}} +(\hat{x}\to -\hat{x}),
\label{qq}\\
\tilde{\kappa}_{qq}(\hat{x},\hat{\xi}) 
&=& \frac{3}{2(1-\hat{x})}+\frac{\hat{x}^2+1-2\hat{\xi}^2}{(1-\hat{\xi}^2)(1-\hat{x})} \ln \frac{\hat{x}-1}{\hat{x}} - \frac{(\hat{x}-\hat{\xi})(1+\hat{x}^2+2\hat{x} \hat{\xi})}{(1-\hat{x}^2) (1-\hat{\xi}^2)} \ln \frac{\hat{x}-\hat{\xi}}{\hat{x}} -(\hat{x}\to -\hat{x}), \label{qqtilde}
\\
\kappa_{qg}(\hat{x},\hat{\xi})
&=&  \frac{1-2\hat{x}+2\hat{x}^2-\hat{\xi}^2}{(1-\hat{\xi}^2)^2}\ln \frac{\hat{x}-1}{\hat{x}} +\frac{(\hat{x}-\hat{\xi})(1-2\hat{\xi} \hat{x} -\hat{\xi}^2)}{\hat{\xi} (1-\hat{\xi}^2)^2}\ln \frac{\hat{x}-\hat{\xi}}{\hat{x}} +(\hat{x}\to -\hat{x}), \label{gq}\\
\tilde{\kappa}_{qg}(\hat{x},\hat{\xi}) 
 &=& \frac{2\hat{x}-1 -\hat{\xi}^2}{(1-\hat{\xi}^2)^2}\ln \frac{\hat{x}-1}{\hat{x}}  -2\frac{\hat{x}-\hat{\xi}}{(1-\hat{\xi}^2)^2}\ln \frac{\hat{x}-\hat{\xi}}{\hat{x}} -(\hat{x}\to -\hat{x}) , \label{qgtilde}
\eeq
where as before $\hat{x} =\frac{x_B}{x}$, $\hat{\xi}=\frac{\xi}{x}$. Note that $\hat{x},\hat{\xi}$ are always positive because we restricted to $0<x<1$ in (\ref{sym}) and (\ref{asym}). 
Also, an infinitesimal, negative imaginary part is understood in $\hat{x}$, namely, $\hat{x}\to \hat{x}-i\epsilon$  and  
\beq
\ln (\hat{x}-1) = \ln (1-\hat{x}) -i\pi.
\eeq

\pagebreak

\subsection{Coefficient functions}

The `coefficient functions' in (\ref{log1}), (\ref{log2}) are given as follows: 
\begin{align}
\delta C_1^q(\hat{x},\hat{\xi}) &= -\dfrac{ \bigg ( \dfrac{Q^2}{-l^2}\bigg )^{\epsilon_{\IR}}}{\epsilon_{\IR}^2 (1-\hat{x})} - \dfrac{3 \bigg ( \dfrac{Q^2}{-l^2}\bigg )^{\epsilon_\IR}}{2\epsilon_\IR (1-\hat{x})} + \dfrac{1-2\hat{x}-2\hat{x}^2+3\hat{\xi}^2}{2(1-\hat{x}) (1-\hat{\xi}^2)} \ln \dfrac{\hat{x}-1}{\hat{x}} + \dfrac{(\hat{x}-\hat{\xi}) (-1+\hat{x}^2+3\hat{x} \hat{\xi} +3 \hat{\xi}^2)}{(1-\hat{x}^2) (1-\hat{\xi}^2) \hat{\xi}} \ln \dfrac{\hat{x}-\hat{\xi}}{\hat{x}} \nonumber \\[0.2cm]
& + \dfrac{1+\hat{x}^2-2\hat{\xi}^2}{2(1-\hat{x})(1-\hat{\xi}^2)} \ln^2 \dfrac{\hat{x}-1}{\hat{x}} + \dfrac{\hat{x}}{ 2(1-\hat{\xi}^2) \hat{\xi}} \ln^2 \dfrac{\hat{x}-\hat{\xi}}{\hat{x}} + \dfrac{-1-\hat{x}^2 +2\hat{\xi}^2}{2(1-\hat{x}^2) (1-\hat{\xi}^2)} \ln \dfrac{\hat{x}-\hat{\xi}}{\hat{x}} \ln \dfrac{\hat{x}+\hat{\xi}}{\hat{x}} + \dfrac{\pi^2 -54}{12(1-\hat{x})} \nonumber \\[0.2cm]
& + \dfrac{\hat{x}}{(1-\hat{\xi}^2) \hat{\xi}} \textrm{Li}_2 \dfrac{-2\hat{\xi}}{\hat{x}-\hat{\xi}} + \dfrac{1+\hat{x}^2 -2\hat{\xi}^2}{(1-\hat{x}) (1-\hat{\xi}^2)} \left( \textrm{Li}_2 \dfrac{1-\hat{\xi}}{1-\hat{x}} +  \textrm{Li}_2 \dfrac{1+\hat{\xi}}{1-\hat{x}}\right) \, + (\hat{x} \to -\hat{x}),
\label{cq}
\end{align}
\begin{align}
\delta \tilde{C}_1^q(\hat{x},\hat{\xi})  & = -\dfrac{ \bigg ( \dfrac{Q^2}{-l^2}\bigg )^{\epsilon_\IR}}{\epsilon_\IR^2 (1-\hat{x})} - \dfrac{3 \bigg ( \dfrac{Q^2}{-l^2}\bigg )^{\epsilon_\IR}}{2\epsilon_{\IR} (1-\hat{x})} + \dfrac{-1+2\hat{x}-4\hat{x}^2 +3\hat{\xi}^2}{2(1-\hat{x}) (1-\hat{\xi}^2)} \ln \dfrac{\hat{x}-1}{\hat{x}} + \dfrac{(\hat{x}-\hat{\xi})(1+2\hat{x}^2 +3\hat{x} \hat{\xi})}{(1-\hat{x}^2)(1-\hat{\xi}^2)} \ln \dfrac{\hat{x}-\hat{\xi}}{\hat{x}} \nonumber \\[0.2cm]
& + \dfrac{1+\hat{x}^2 -2\hat{\xi}^2}{2(1-\hat{x}) (1-\hat{\xi}^2)} \ln^2 \dfrac{\hat{x}-1}{\hat{x}} + \dfrac{\hat{\xi}}{2(1-\hat{\xi}^2)} \ln^2 \dfrac{\hat{x}-\hat{\xi}}{\hat{x}} - \dfrac{\hat{x} (1+\hat{x}^2 -2\hat{\xi}^2)}{2(1-\hat{x}^2) (1-\hat{\xi}^2)} \ln \dfrac{\hat{x}-\hat{\xi}}{\hat{x}} \ln \dfrac{\hat{x}+\hat{\xi}}{\hat{x}} + \dfrac{\pi^2 -54}{12(1-\hat{x})} \nonumber \\[0.2cm]
& +\dfrac{\hat{\xi}}{1-\hat{\xi}^2} \textrm{Li}_2 \dfrac{-2\hat{\xi}}{\hat{x}-\hat{\xi}} + \dfrac{1+\hat{x}^2 - 2\hat{\xi}^2}{(1-\hat{x}) (1-\hat{\xi}^2)} \left( \textrm{Li}_2 \dfrac{1-\hat{\xi}}{1-\hat{x}} + \textrm{Li}_2 \dfrac{1+\hat{\xi}}{1-\hat{x}}\right)  -(\hat{x}\to -\hat{x}),
\label{cq2} 
\\[0.5cm]
\delta C_1^g(\hat{x},\hat{\xi}) & =  -\dfrac{ 1-2\hat{x}+2\hat{x}^2 -\hat{\xi}^2}{(1-\xi^2)^2} \ln \dfrac{\hat{x}-1}{\hat{x}}  + \dfrac{1-2\hat{x}+2\hat{x}^2-\hat{\xi}^2}{2(1-\hat{\xi}^2)^2} \ln^2 \dfrac{\hat{x}-1}{\hat{x}} \nn
 &- \dfrac{(\hat{x}-\hat{\xi})(1-2\hat{x}\hat{\xi}-\hat{\xi}^2)}{\hat{\xi}(1-\hat{\xi}^2)^2} \ln \dfrac{\hat{x}-\hat{\xi}}{\hat{x}}  + \dfrac{\hat{x}(1+\hat{\xi}^2)}{2\xi(1-\hat{\xi}^2)^2 } \ln^2 \dfrac{\hat{x}-\hat{\xi}}{\hat{x}} - \dfrac{1+2\hat{x}^2 -\hat{\xi}^2}{2(1-\hat{\xi}^2)^2} \ln \dfrac{\hat{x}-\hat{\xi}}{\hat{x}} \ln \dfrac{\hat{x}+\hat{\xi}}{\hat{x}} \nonumber \\[0.2cm]
 & + \dfrac{\hat{x}(1+\hat{\xi}^2)}{\hat{\xi}(1-\hat{\xi}^2)^2} \textrm{Li}_2 \dfrac{-2\hat{\xi}}{\hat{x}-\hat{\xi}} + \dfrac{1-2\hat{x}+2\hat{x}^2-\hat{\xi}^2}{(1-\hat{\xi}^2)^2} \left(\textrm{Li}_2 \dfrac{1-\hat{\xi}}{1-\hat{x}} +  \textrm{Li}_2 \dfrac{1+\hat{\xi}}{1-\hat{x}}\right) \, + (\hat{x} \to -\hat{x}) , \label{cg} \\[0.5cm]
\delta\tilde{C}_1^g(\hat{x},\hat{\xi}) & =- \dfrac{2\hat{x}-1-\hat{\xi}^2}{(1-\hat{\xi}^2)^2} \ln \dfrac{\hat{x}-1}{\hat{x}} + \dfrac{2\hat{x}-1-\hat{\xi}^2}{2(1-\hat{\xi}^2)^2} \ln^2 \dfrac{\hat{x}-1}{\hat{x}}  \nonumber \\[0.2cm]
 & + 2\dfrac{\hat{x}-\hat{\xi}}{(1-\hat{\xi}^2)^2} \ln \dfrac{\hat{x}-\hat{\xi}}{\hat{x}} + \dfrac{ \hat{\xi}}{(1-\hat{\xi}^2)^2} \ln^2 \dfrac{\hat{x}-\hat{\xi}}{\hat{x}} - \dfrac{\hat{x}}{(1-\hat{\xi}^2)^2} \ln \dfrac{\hat{x}-\hat{\xi}}{\hat{x}} \ln \dfrac{\hat{x}+\hat{\xi}}{\hat{x}} \nonumber \\[0.2cm]
 & + \dfrac{ 2\hat{\xi}}{(1-\hat{\xi}^2)^2} \textrm{Li}_2 \dfrac{-2\hat{\xi}}{\hat{x}-\hat{\xi}} + \dfrac{2\hat{x}-1-\hat{\xi}^2}{(1-\hat{\xi}^2)^2} \left(\textrm{Li}_2 \dfrac{1-\hat{\xi}}{1-\hat{x}} +  \textrm{Li}_2 \dfrac{1+\hat{\xi}}{1-\hat{x}}\right) \, - (\hat{x} \to -\hat{x}), \label{cg2}
\end{align}
where ${\rm Li}_2$ is the dilogarithm function.  
As mentioned before, in the quark channel, we find single $1/\epsilon_\IR$ and double $1/\epsilon_\IR^2$ infrared poles. 
Note that, since $\epsilon_\IR<0$,  $({Q^{2}/}{-l^2})^{\epsilon_\IR} \to 0$ if one takes the $l^2\to 0$ limit first. However, if one keeps $l^2$ finite and expands in $\epsilon_\IR$,  the first term $\frac{3}{2(1-\hat{x})}$ in (\ref{qq}) and (\ref{qqtilde}) gets  canceled. We discuss below how  these problematic terms eventually disappear.  As we also mentioned, the result (\ref{cq2}) is independent of the scheme for $\gamma_5$.

\subsection{Anomaly pole terms}

The coefficients of the `anomaly poles' $1/t$ in (\ref{sym}) and (\ref{asym}) are found to be
\beq
A(x,x_B,\xi) &=& \frac{2T_R}{x}\left(1+ \frac{ \hat{x}(1-\hat{x}) \ln \dfrac{\hat{x}-1}{\hat{x}} + \hat{x}(\hat{x}-\hat{\xi}) \ln \dfrac{\hat{x}-\hat{\xi}}{\hat{x}} +(\hat{x}\to -\hat{x}) }{1-\hat{\xi}^2} \right), \label{a1}  \\
\tilde{A}(x,x_B,\xi)&=&\frac{8T_R}{x} \frac{(1-\hat{x})\ln\frac{\hat{x}-1}{x}+(\hat{x}-\hat{\xi})\ln\frac{\hat{x}-\hat{\xi}}{\hat{x}}
-(\hat{x}\to -\hat{x})}{1-\hat{\xi}^2}.\label{a2}
\eeq
The imaginary parts of these expressions  agree with our results in \cite{Bhattacharya:2022xxw}. Clearly,  there are $1/t$ poles also in the real part of the Compton amplitude.\footnote{In Ref.~\cite{Bhattacharya:2022xxw} we 
  computed only the imaginary part of the Compton amplitude by directly extracting the discontinuity across the variables $s=(p+q)^2$ and $q^2_2$. In this paper, we compute the full amplitude. We have checked that  the imaginary parts of (\ref{a1}),  (\ref{a2}) and all the other results in this paper are  consistent with the corresponding  results in~\cite{Bhattacharya:2022xxw}. 
  } 
While (\ref{a1}) and (\ref{a2}) look unfamiliar and complicated, remarkably  the $x$-integrals in (\ref{sym}) and (\ref{asym}) can be exactly rewritten in the following form
\beq
&&\int_0^1dx A(x,x_B,\xi) {\cal F}(x,\xi,t) \nn 
&& =2T_R\int_0^1 dx C_0(x,x_B) \left[\int_x^1 \frac{dx'}{x'}K\left(\frac{x}{x'},\frac{\xi}{x'}\right){\cal F}(x',\xi,t) -\theta(\xi-x)\int_0^1 \frac{dx'}{x'} L\left(\frac{x}{x'},\frac{\xi}{x'}\right){\cal F}(x',\xi,t)\right]  \nn
&& \equiv 2T_R\int_0^1 dx C_0(x,x_B) C^{\rm anom}\otimes {\cal F}(x,\xi,t),
\label{white}
\eeq
\beq
&&\int_0^1 dx \tilde{A}(x,x_B,\xi)\tilde{\cal F}(x,\xi,t) \nn 
&& =2T_R\int_0^1 dx \tilde{C}_0(x,x_B) \left[\int_x^1 \frac{dx'}{x'}\tilde{K}\left(\frac{x}{x'},\frac{\xi}{x'}\right)\tilde{\cal F}(x',\xi,t) -\theta(\xi-x)\int_0^1 \frac{dx'}{x'} \tilde{L}\left(\frac{x}{x'},\frac{\xi}{x'}\right)\tilde{\cal F}(x',\xi,t)\right]  \nn
&& \equiv 2T_R\int_0^1 dx \tilde{C}_0(x,x_B) \tilde{C}^{\rm anom}\otimes\tilde{\cal F}(x,\xi,t), \label{white2}
\eeq 
 where 
\beq
&&K (x,\xi)  
=\frac{x(1-x)}{1-\xi^2} \, , \qquad L(x,\xi) = \frac{x(\xi-x)}{1-\xi^2} \, , \label{KL1} \\
&& \tilde{K}(x,\xi)=\frac{4(1-x)}{1-\xi^2}, \qquad \tilde{L}(x,\xi)=\frac{4(\xi-x)}{1-\xi^2}. \label{KL2}
\eeq
That is, the leading-order kernels $C_0$ and $\tilde{C}_0$ can be factored out. The resulting convolution $C^{\rm anom}\otimes {\cal F}$ agrees with what was anticipated in \cite{Bhattacharya:2022xxw} following the general argument in \cite{White:2001pu} where actually the same integral (\ref{white}) can be found. We now have the corresponding result with $\xi \neq 0$ in the polarized sector. As mentioned already in \cite{Bhattacharya:2022xxw}, the two terms in $C^{\rm anom}$ and $\bar{C}^{\rm anom}$ come from the first and third diagrams in Fig.~\ref{fig1}. The latter is nonzero only when the outgoing photon becomes timelike, $q_2^2>0$, see (\ref{timelike}). 

The identities (\ref{white}), (\ref{white2})  guarantee that, if the $1/t$ poles are cancelled in the imaginary part of the Compton amplitude  \cite{Bhattacharya:2022xxw}, the same cancellation automatically occurs  in the real part as well. 


\section{GPD at one loop}
\label{gpd1loop}

We have seen that the Compton scattering amplitudes at one loop contain three types of singular behaviors: (i) logarithms $\ln (-t)$, (ii) anomaly poles $1/t$, (iii) single $1/\epsilon$ and double  $1/\epsilon^2$ infrared poles  (only in the quark channel). The  logarithms are as expected, but the other two are unusual and potentially cause problems with factorization. We now demonstrate that all these singular structures can be absorbed into the quark GPDs in the leading-order terms of (\ref{sym}) and (\ref{asym}). Specifically, we compute the unpolarized and polarized quark 
GPDs\footnote{The variables $x$ and $\xi$ in this section (and also in the appendix) should better be written as $\hat{x}$ and $\hat{\xi}$ to be more consistent with the notation in  the  previous sections. We however abbreviate $\hat{x},\hat{\xi}\to x,\xi$ for simplicity.}
\beq
f_q(x,\xi,t)&=&\int \frac{dz^-}{ 4\pi}e^{ixP^+z^-}\langle p_2|\bar{q}(-z/2)W\gamma^+q(z^-/2)|p_1\rangle, \label{unpolq}\\
\tilde{f}_q(x,\xi,t)&=& \int \frac{dz^-}{4\pi}e^{ixP^+z^-}\langle p_2|\bar{q}(-z/2)W\gamma^+\gamma_5q(z^-/2)|p_1\rangle, \label{polq}
\eeq
to one loop for on-shell $p_1^2=p_2^2=0$ quark and gluon external states, keeping $t=(p_2-p_1)^2\neq 0$. We need to separately consider the  DGLAP region $0<\xi < x\le 1$ and  the Efremov-Radyushkin-Brodsky-Lepage (ERBL) region $0<x<\xi$ \cite{Diehl:2003ny}. We work in $d=4-2\epsilon$ dimensions to regularize the UV divergences and any leftover IR divergences. 
As before, they are distinguished by $1/\epsilon_{\UV}$ and $1/\epsilon_{\IR}$, respectively. The $\overline{\rm MS}$ scale is denoted by $\tilde{\mu}^2=4\pi e^{-\gamma_E}\mu^2$.

\subsection{Quark matrix element}

The divergent part of the quark matrix element in the DGLAP region $\xi<x<1$ is, omitting the common prefactor $\frac{\alpha_sC_F}{ 4\pi p^+}\bar{u}(p_2)\gamma^+u(p_1)$ or   $\frac{\alpha_sC_F}{4\pi p^+}\bar{u}(p_2)\gamma^+\gamma_5u(p_1)$, 
\beq
\frac{\left(\frac{\tilde{\mu}^2}{-l^2}\right)^\epsilon}{\epsilon_{\UV}} \left[\frac{1+x^2-2\xi^2}{(1-\xi^2)(1-x)_+} +\left(\frac{3}{2}-\ln (1-\xi^2)\right)\delta(1-x)\right] - \delta(1-x)\left( \frac{1}{\epsilon^2_{\IR}}+ \frac{3}{2}\frac{1}{\epsilon_{\IR}}\right)\left(\frac{\tilde{\mu}^2}{-l^2}\right)^\epsilon.\label{dg}
\eeq
The derivation is outlined in Appendix A. 
The  same result holds for both the unpolarized and polarized GPDs. Note the double and single infrared poles. Similar `Sudakov' double poles have been previously encountered in the one-loop gluon matrix element of the twist-four GPDs (\ref{four}) and (\ref{two}) \cite{Radyushkin:2021fel,Radyushkin:2022qvt}.  After convoluting with $C_0$ and $\tilde{C}_0$, which can be done trivially using the delta function $\delta(1-x)$, they exactly match  the double and single poles in (\ref{cq}) and (\ref{cq2}). In other words, the poles in (\ref{cq}) and (\ref{cq2}) can be absorbed into the GPDs.  The finite terms are given by 
\beq
 - \frac{1+x^2-2\xi^2}{1-\xi^2} \left(\frac{\ln\frac{(1-x)^2}{1-\xi^2}}{1-x}\right)_+ - \frac{1-x}{1-\xi^2} -\frac{1}{2}\delta(1-x)\left( \ln^2(1-\xi^2) -\frac{\pi^2}{6}\right) , \label{finunpol}
\eeq
in the unpolarized case. In the polarized case, interestingly the result depends on the scheme for $\gamma_5$ in contrast to what we observed with the coefficient function (\ref{cq2}). If one uses the `fully anticommuting $\gamma_5$,' the result is the same as (\ref{finunpol}). If, on the other hand, one uses the HVBM scheme, one finds an additional term 
\beq
4\frac{1-x}{1-\xi^2}. \label{extra}
\eeq
The $\mathcal{O}(\epsilon)$ difference between the two schemes now leaves a finite contribution because of the presence of the UV pole $1/\epsilon_\UV$.  

In the ERBL region, the divergent terms read 
\beq
\frac{\left(\frac{\tilde{\mu}^2}{-l^2}\right)^\epsilon}{ \epsilon_{\UV}}\frac{x+\xi}{(1+\xi)}\left(\frac{1}{1-x}+\frac{1}{2\xi}\right), \label{dge}
\eeq
in both the unpolarized and polarized cases. 
The finite terms are rather cumbersome. In the unpolarized case, we find 
\beq
&&\frac{1}{2\xi(1-x) (1-\xi^2)} \Biggl[(x-1)(x+\xi)(\xi+1)\ln (\xi^2-x^2)+2\xi x^2\ln \frac{1+\xi}{(x+\xi)(1-x)}+2\xi^3\ln \frac{x+\xi}{\xi-x} \nn
&&\qquad +2\xi\ln \frac{(1+\xi)(\xi-x)}{1-x}+4\xi^3\ln \frac{1-x}{1+\xi} +2(1-x)(x+\xi^2)\ln (2\xi)\Biggr] - \frac{x+\xi}{2\xi(1+\xi)},
\label{e:ERBl_fin}
\eeq
and in the polarized case there is an additional term 
\beq
4\frac{x+\xi}{2\xi (1+\xi)},
\eeq
in the HVBM scheme.  

The coefficients of the UV pole in (\ref{dg}) and (\ref{dge}) constitute the $q\to q$ evolution kernel of the GPDs. 
After expanding $\left(\frac{\tilde{\mu}^2}{-l^2}\right)^\epsilon=1+\epsilon \ln \frac{\tilde{\mu}^2}{-l^2}$ and 
convoluting with $C_0$ and $\tilde{C}_0$, we recover (\ref{qq}) and (\ref{qqtilde}). In other words, the logarithmic terms in (\ref{log1})  can be absorbed into the GPDs, as expected. The same comment applies to the $g\to q$ evolution kernel below.

\subsection{Gluon matrix element, unpolarized}

For the gluon matrix elements  $\langle p_2\epsilon_2|...|p_1\epsilon_1\rangle$ of the quark GPD (\ref{unpolq}), (\ref{polq}), we find it  convenient  to work in the light-cone gauge $\epsilon_1^+=\epsilon_2^+=0$.    
 The result for the unpolarized GPD in the DGLAP region is
\beq
f_q&=&\frac{\alpha_sT_R}{2\pi} \Biggl[-(1-\xi^2)\epsilon_1\cdot \epsilon_2^* \left( \frac{ \left(\frac{\tilde{\mu^2}}{-l^2}\right)^\epsilon}{\epsilon_{\rm UV}} \frac{2x^2-2x+1-\xi^2}{(1-\xi^2)^2} - \frac{(2x^2-2x+1-\xi^2) \ln \frac{(1-x)^2}{1-\xi^2}+2x(1-x)}{(1-\xi^2)^2}\right) \nn 
&& \qquad\qquad  - \frac{4}{l^2} \frac{x(1-x)}{1-\xi^2} \left(\epsilon_1\cdot l \epsilon_2^*\cdot l -\frac{l^2}{2}\epsilon_1\cdot\epsilon_2^*\right) \Biggr], \label{glunp}
\eeq
where we have factored out the structure that represents the twist-two GPD, see (\ref{matr}). 
Note the anomaly pole $1/l^2$ which was absent in the quark matrix elements. 
Its coefficient matches $K$ in (\ref{KL1}). (The factor of 4 is from (\ref{four}).)

In the ERBL region, we find,  omitting the prefactor $\frac{\alpha_sT_R}{2\pi}$, 
\beq
 -(1-\xi^2)\epsilon_1\cdot \epsilon_2^*\frac{\left(\frac{\tilde{\mu}^2}{-l^2}\right)^\epsilon}{\epsilon_{\UV}}\frac{(x+\xi)(1+\xi-2x)}{2\xi(1+\xi)(1-\xi^2)}  -  \left(\epsilon_1\cdot l \epsilon_2^*\cdot l -\frac{l^2}{2}\epsilon_1\cdot\epsilon_2^*\right)\frac{1}{l^2}\frac{2x(x+\xi)}{\xi(1+\xi)} -(x\to -x),
\eeq 
The first term comes from the same ladder diagram as in (\ref{glunp}). The $x\to -x$ term comes from a diagram in which the gluon legs are crossed. The latter contributes only in the ERBL region.  The finite terms read, including the $x\to -x$ contribution,
\beq
-(1-\xi^2)\epsilon_1\cdot \epsilon_2^* \frac{1}{\xi(1-\xi^2)^2}\Biggl[ - x(1+\xi^2) \ln \frac{\xi^2-x^2}{4}-\xi(1+2x^2-\xi^2) \ln \frac{(x+\xi)(1-x)}{(\xi-x)(1+x)}\nn +2x \Bigl(\xi \ln (1-x^2) -2\xi\ln(1+\xi)+\xi^2-\xi+(1+\xi^2)\ln \xi\Bigr)\Biggr]. \label{finite2}
\eeq
The coefficient of the anomaly pole is thus 
\beq
\frac{1}{l^2}\left(\frac{2x(x+\xi)}{\xi(1+\xi)}-\frac{2x(x-\xi)}{\xi(1+\xi)}\right) = \frac{4}{l^2}\frac{x}{1+\xi}.
\eeq
 This agrees with 
\beq
K(x,\xi)-L(x,\xi) 
= \frac{x}{1+\xi},
\eeq
which is the relevant linear combination in the ERBL region, see  (\ref{white}). These results, together with the observation (\ref{white}), show that the $1/t$ pole  in (\ref{sym}) can be absorbed into the unpolarized quark GPDs $H_q,\, E_q$ in the leading order as a part of the infrared subtraction procedure.

\subsection{Gluon matrix element, polarized}

In the polarized case (\ref{polq}), we find 
\beq
\tilde{f}_q=\frac{\alpha_sT_R}{2\pi} \left[(1-\xi^2)i\epsilon^{+p\epsilon_2^* \epsilon_1} \left(\frac{2x-1-\xi^2}{(1-\xi^2)^2}\left(\frac{\left(\frac{\tilde{\mu}^2}{-l^2}\right)^\epsilon}{\epsilon_{\UV}} -\ln \frac{(1-x)^2}{1-\xi^2}\right) -2\frac{1-x}{(1-\xi^2)^2}\right) + \frac{2il^+\epsilon^{\epsilon_1 \epsilon_2^* lp}}{l^2} \frac{1-x}{1-\xi^2}\right], \label{polg} 
\eeq
in the DGLAP region. 
A simplified version of this result (with a different UV prescription) was already reported in  \cite{Bhattacharya:2022xxw}. The coefficient of the pole agrees with $\tilde{K}$ in (\ref{KL2}).  Note that the pole is proportional to $l^{\mu=+}$, meaning that it contributes to a shift in the GPD $\tilde{E}_q$.  
In the ERBL region, the singular terms are, omitting the prefactor $\frac{\alpha_sT_R}{2\pi}$, 
\beq
&&\frac{x+\xi}{2\xi(1+\xi)} \left((1-\xi^2)i\epsilon^{+p\epsilon_2^* \epsilon_1}\frac{\left(\frac{\tilde{\mu}^2}{-l^2}\right)^\epsilon}{\epsilon_{\UV}}\frac{-1}{1+\xi}  + \frac{2il^+\epsilon^{\epsilon_1 \epsilon_2^* lp}}{l^2}\right)+(x \to -x)\nn 
&& \quad =(1-\xi^2)i\epsilon^{+p\epsilon_2^* \epsilon_1}\frac{\left(\frac{\tilde{\mu}^2}{-l^2}\right)^\epsilon}{\epsilon_{\UV}} \frac{-1}{(1+\xi)^2} + \frac{2il^+\epsilon^{\epsilon_1 \epsilon_2^* lp}}{l^2} \frac{1}{1+\xi}, 
\eeq
where again the $x\to -x$ term comes from the crossed-leg diagram. 
 The finite terms are, including the $x\to -x$ contribution,
\beq
&& (1-\xi^2)i\epsilon^{+p\epsilon_2^* \epsilon_1}\frac{1}{(1-\xi^2)^2}\Biggl[ -2\xi \ln (\xi^2-x^2)+(1+\xi^2)\ln (1-x^2)-2x \ln \frac{(1-x)(x+\xi)}{(1+x)(\xi-x)} \nn 
&& \qquad \qquad  -2(1+\xi^2)\ln (1+\xi)+4\xi \ln (2\xi) +2\xi -2  \Biggr]. \label{finite3}
\eeq 
The coefficient of the UV pole $\frac{-1}{(1+\xi)^2}$ is the correct evolution kernel in the ERBL region as can be seen by taking the imaginary part 
of (\ref{qgtilde}):
\beq
\frac{2x-1-\xi^2}{(1-\xi^2)^2}-\frac{2(x-\xi)}{(1-\xi^2)^2}= \frac{-1}{(1+\xi)^2}.
\eeq
Again the coefficient of the anomaly pole agrees with 
\beq
\tilde{K}(x,\xi)-\tilde{L}(x,\xi) = \frac{4}{1+\xi},
\eeq
from (\ref{KL2}). 
With the help of (\ref{white2}), we can absorb the $1/t$ pole in (\ref{asym})  into the polarized quark GPD $\tilde{E}_q$.

\subsection{Relation to the $\overline{\rm MS}$ scheme}

We  have thus shown that all the singular structures $1/t$, $\ln (-t)$, $1/\epsilon_{\rm IR}$ and $1/\epsilon^2_{\rm IR}$ in the `unsubtracted' expressions (\ref{sym}) and (\ref{asym})  can be systematically absorbed into the twist-two GPDs in the leading order. Since the matrix elements (\ref{unpolq}) and (\ref{polq})  contain non-singular terms,  one might choose to perform this subtraction also for the finite terms (\ref{cq})-(\ref{cg2}).
An interesting question then arises as to whether, after such a subtraction, (\ref{cq})-(\ref{cg2}) reduce to the known coefficient functions in the $\overline{\rm MS}$ scheme.\footnote{We thank Vladimir Braun for raising this question.} Here we partially address this question by explicitly performing the subtraction in the imaginary part of the Compton amplitude.

Let us first consider the DGLAP region $0<\xi<x$. For simplicity, we assume $x<1$ to avoid the delta function $\delta(1-x)$.  The imaginary part of (\ref{cq})  is 
\beq
\frac{1-2x-2x^2+3\xi^2}{2(1-x)(1-\xi^2)} + \frac{1+x^2-2\xi^2}{(1-x)(1-\xi^2)}\ln \frac{1-\xi^2}{x(1-x)} , \label{c1}
\eeq
where we used 
\beq
{\rm Im}\, {\rm Li}_2\frac{1\pm \xi}{1-x+i\epsilon} = -\pi \ln \frac{1\pm \xi}{1-x}.
\eeq
On the other hand, the finite terms in the unpolarized quark GPD are, from (\ref{finunpol}),
\beq
-\frac{1+x^2-2\xi^2}{(1-x)(1-\xi^2)}\ln \frac{(1-x)^2}{1-\xi^2} -\frac{1-x}{1-\xi^2}. \label{as}
\eeq
The convolution with the leading-order kernel (\ref{leading}) is trivial for the imaginary part since ${\rm Im}\, C_0\propto  \delta(1-x)$. We just need to subtract (\ref{as}) from (\ref{c1}) to obtain 
\beq
\frac{1+x^2-2\xi^2}{(1-x)(1-\xi^2)}\ln \frac{1-x}{x} + \frac{3(1-2x+\xi^2)}{2(1-x)(1-\xi^2)}
\to 2+x-\frac{3}{2(1-x)}
 + \frac{1+x^2}{1-x}\ln \frac{1-x}{x} + 1-x,
\eeq
where we have
set $\xi=0$ on the right-hand side. 
This agrees with the imaginary part of the $q\to q$ coefficient function in the $\overline{\rm MS}$ scheme \cite{Ji:1997nk,Belitsky:1997rh,Mankiewicz:1997bk,Pire:2011st,Bertone:2022frx}. In particular,  the right-hand side is the familiar $q\to q$ coefficient function for the $F_1$ structure function in DIS  \cite{Altarelli:1979ub} for $x<1$.
Similarly, the imaginary part of (\ref{cq2}) reads 
\beq
\frac{-1+2x-4x^2+3\xi^2}{2(1-x)(1-\xi^2)} + \frac{1+x^2-2\xi^2}{(1-x)(1-\xi^2)}\ln \frac{1-\xi^2}{x(1-x)} .
\eeq
The finite terms in the polarized quark PDF depend on the scheme adopted for the treatment of $\gamma_5$. In the HVBM scheme, we find from (\ref{finunpol}) and  (\ref{extra}), 
\beq
-\frac{1+x^2-2\xi^2}{(1-x)(1-\xi^2)}\ln \frac{(1-x)^2}{1-\xi^2} +3\frac{1-x}{1-\xi^2}.
\eeq
After the subtraction, 
\beq
\frac{-7+14x-10x^2+3\xi^2}{2(1-x)(1-\xi^2)} + \frac{1+x^2-2\xi^2}{(1-x)(1-\xi^2)}\ln \frac{1-x}{x} \to  2+x-\frac{3}{2(1-x)} + \frac{1+x^2}{1-x}\ln \frac{1-x}{x} -4( 1-x), \label{hv}
\eeq
in agreement with the $q\to q$ coefficient function for the $g_1$ structure function in the HVBM prescription. 
As was discussed in Refs.~\cite{Stratmann:1995fn,Vogelsang:1996im}, it is necessary to subtract this 
term in order to avoid a conflict with helicity conservation and with the known first-order correction to the
Bjorken sum rule. Incidentally, in the present case, the result obtained after this finite subtraction coincides 
with that found for a fully anticommuting $\gamma_5$. Either way, instead of~(\ref{hv}) the correct result becomes 
\beq
\frac{1-2x-2x^2+3\xi^2}{2(1-x)(1-\xi^2)}  + \frac{1+x^2-2\xi^2}{(1-x)(1-\xi^2)}\ln \frac{1-x}{x} \to  2+x-\frac{3}{2(1-x)} + \frac{1+x^2}{1-x}\ln \frac{1-x}{x} .
\eeq

As for the $g\to q$ coefficients,   the imaginary part of (\ref{cg}) is 
\beq
\frac{1-2x+2x^2-\xi^2}{(1-\xi^2)^2}\left(\ln\frac{1-\xi^2}{x(1-x)}-1\right) .
\eeq
From this, we subtract the finite terms in  (\ref{glunp}),
\beq
-\frac{1-2x+ 2x^2-\xi^2}{(1-\xi^2)^2}\ln \frac{(1-x)^2}{1-\xi^2} -\frac{2x(1-x)}{(1-\xi^2)^2},
\eeq
to obtain  
\beq
\frac{1-2x+2x^2-\xi^2}{(1-\xi^2)^2}\ln\frac{1-x}{x} + \frac{-1+4x-4x^2+\xi^2}{(1-\xi^2)^2} \to (1-2x+2x^2)\left(\ln\frac{1-x}{x}-1\right) + 2x(1-x). \label{f1g}
\eeq
This agrees with the $g\to q$ coefficient function for the $F_1$ structure function in the $\overline{\rm MS}$ scheme. 
Finally, the imaginary part of (\ref{cg2}) is 
\beq
\frac{2x-1-\xi^2}{(1-\xi^2)^2}\left(\ln \frac{1-\xi^2}{x(1-x)}-1\right).
\eeq 
Subtracting from this the finite terms in (\ref{polg}),
\beq
-\frac{2x-1-\xi^2}{(1-\xi^2)^2} \ln \frac{(1-x)^2}{1-\xi^2}-\frac{2(1-x)}{(1-\xi^2)^2}, \label{uv}
\eeq 
we find 
\beq
\frac{2x-1-\xi^2}{(1-\xi^2)^2} \ln \frac{1-x}{x}+\frac{3-4x+\xi^2}{(1-\xi^2)^2} \to  (2x-1)\left(\ln \frac{1-x}{x} -1\right) +2(1-x),
\label{last}
\eeq
in agreement with the $\overline{\rm MS}$
$g\to q$ coefficient function for the $g_1$ structure function. It is interesting to recall that the last term $2(1-x)\otimes \Delta G(x)$ in (\ref{last})  caused a lot  of discussion (see, e.g., \cite{Bodwin:1989nz,Vogelsang:1990ug}) in the wake of the proton `spin crisis' in the late 80s. 
 In the standard $\overline{\rm MS}$ calculation in the forward limit $t=0$, this term arises from the IR region of the loop diagram, and therefore does not  seem to qualify as a part of the `hard' coefficient. 
 In our calculation of the Compton amplitude, this term is replaced by the pole term $\frac{1}{l^2}(1-x)\otimes \tilde{\cal F}(x)$ and gets absorbed into the GPD $\tilde{E}_q$. Nevertheless,   the $2(1-x)$ term is restored after the subtraction because the polarized GPD (\ref{polq}) generates it from the UV region of the loop momentum. Therefore, even though the final result (\ref{last}) is the same, from our perspective the term $2(1-x)$ is legitimately considered  a `hard' contribution. 
 
We have further performed the subtraction of the constant terms (\ref{e:ERBl_fin}), (\ref{finite2}) and (\ref{finite3}) in the ERBL region $x<\xi$ from the imaginary part of (\ref{cq})-(\ref{cg2}) and observed consistent agreement with the $\overline{\rm MS}$ coefficient functions \cite{Belitsky:2005qn}. We have thus partially verified that the ``off-forward" regularization is equivalent to the $\overline{\rm MS}$ scheme after the subtraction of finite terms is made. Extending this to the  real part of the Compton amplitude is left for future work. On the other hand, since the treatment of finite terms is a matter of scheme choice, one can choose to subtract only the singular terms. Eqs.~(\ref{cq})-(\ref{cg2}), with the single and double poles omitted, are then the coefficient functions in such a  scheme.  

\section{Imprints of anomalies on GPD}

Let us discuss the implications of our results. Superficially, it may seem as if nothing has happened in the end. After absorbing all the singular terms into the twist-two GPDs, the Compton amplitude will be given by the usual factorized form  with possibly different coefficient functions due to a different scheme choice. 
The common attitude is that one does not care about these singular terms once they have been `discarded'
into a parton distribution, as they will be taken care of by the nonperturbative QCD dynamics. One can also take the view that the $1/t$ poles should disappear in the limit $t\to 0$, because  nonperturbative effects must intervene when $\sqrt{|t|}\sim \Lambda_{\rm QCD}$. However, from the point of view of factorization, technically speaking one is allowed to choose  any infrared regulator that can isolate the collinear divergences, as long as they can be eventually absorbed into parton distributions when the latter are computed with the same IR regulator. In this sense, the use of $t$  is no different from  other regulators such as the current quark mass $m_q$ and the dimensionality $d\neq 4$. One may even argue that it is a more physical scheme, since $t\neq 0$ in actual experiments and naturally cuts off collinear divergences.

Technicalities aside, the real reason we are pursuing the off-forward calculation with nonzero $t$ is that this approach has the potential to uncover  nonperturbative connections between GPDs and QCD anomalies.   Indeed, the very idea that twist-four GPDs are absorbed into twist-two GPDs is quite non-standard and needs to be investigated further, rather than dismissed as a routine infrared subtraction procedure. This is all the more so because, as discussed in \cite{Tarasov:2021yll,Bhattacharya:2022xxw} and elaborated further below, the results we shall get are consistent with what we know about the axial and gravitational form factors which are certain moments of the twist-two GPDs. 

\subsection{Axial and gravitational form factors}

\subsubsection{Isovector axial form factors}

In order to motivate our discussion, let us first recall the familiar example of the isovector axial current $J_{5a}^\alpha = \sum_q \bar{q}\gamma^\alpha \gamma_5 \frac{\tau^a}{2}q$ where $\tau^{a=1,2,3}$ are the Pauli matrices. Its nucleon matrix element is parameterized by the axial form factors,
\beq
\langle P_2|J_{5a}^{\alpha} |P_1\rangle
= \bar{u}(P_2)\left[\gamma^\alpha\gamma_5 F_A(t) +\frac{l^\alpha \gamma_5}{2M}F_P(t)\right]\frac{\tau^a}{2}u(P_1) \, .
\label{factors}
\eeq
In QCD with $n_f=2$ massless flavors,  the current is exactly conserved, $\partial_\alpha J^\alpha_{5a}=0$, due to  chiral symmetry. This imposes a constraint among the form factors 
\beq
2MF_A(t)+\frac{tF_P(t)}{2M}=0.
\eeq
Clearly, $F_P(t)$ has a pole at $t=0$:
\beq
F_P(t) \approx \frac{-4M^2 g_A^{(3)}}{t} \qquad (t\to 0),
\eeq
where $g^{(3)}_A = F_A(0)\approx 1.3$ is the isovector axial coupling constant. The pole is generated by the exchange of the massless pion which is the Nambu-Goldstone boson of spontaneously broken chiral symmetry. This requirement leads to the well-known Goldberger-Treiman relation
\beq
g_A^{(3)}=\frac{f_\pi g_{\pi NN}}{M}, 
\eeq
where $f_\pi$ is the pion decay constant and $g_{\pi NN}$ is the pion-nucleon coupling. 
Now recall that 
$F_P(t)$ is the first moment of the isovector GPD  $\tilde{E}$,
\beq
F_P(t) = \int_{-1}^1 dx \left( \tilde{E}_u(x,\xi,t)- \tilde{E}_d(x,\xi,t)\right).
\eeq
Barring an unlikely possibility that the pole $1/t$ is generated by the $x$-integral, the GPDs themselves hence
have a massless pole at $t=0$:
\beq
\tilde{E}_{u}(x,\xi,t)-\tilde{E}_{d}(x,\xi,t) \sim \theta(\xi-|x|)\frac{g_A^{(3)}}{t} \qquad (t\to 0). \label{pipole}
\eeq
Indeed, such a pole has been discussed in the GPD literature 
(see, e.g., \cite{Penttinen:1999th}) where it has been argued that the pole exists only in the ERBL region $\xi>x$ where GPDs probe the mesonic degrees of freedom inside the nucleon. In actual QCD with massive quarks, the pole is shifted to the physical pion mass,
$\frac{1}{t}\to \frac{1}{t-m_{\pi}^2}$.

\subsubsection{Isoscalar axial form factors}

The story is more complicated for the singlet axial current $J_5^\alpha = \sum_q \bar{q}\gamma^\alpha \gamma_5q$. The associated form factors $g_A$, $g_P$, appearing in the nucleon matrix element via
\beq
\langle P_2|J_5^\alpha |P_1\rangle
= \bar{u}(P_2)\left[\gamma^\alpha\gamma_5 g_A(t) +\frac{l^\alpha \gamma_5}{2M}g_P(t)\right]u(P_1) \,,
\eeq
 are related to the flavor-singlet polarized GPDs as 
\begin{align}
g_A(t) & =\sum_q\int_{-1}^1 dx \tilde{H}_q(x,\xi,t) =\sum_q \int_0^1 dx (\tilde{H}_q(x,\xi,t)+\tilde{H}_{q}(-x,\xi,t)) \, , \label{gaint}\\
g_P(t) & =\sum_q\int_{-1}^1 dx \tilde{E}_q(x,\xi,t) =\sum_q \int_0^1 dx (\tilde{E}_q(x,\xi,t)+\tilde{E}_{q}(-x,\xi,t)) \, . \label{gpint}
\end{align}
In contrast to the isovector current above,  $J_5^\alpha$ is not conserved due to the chiral (U$_A$(1)) anomaly,
\beq
 \partial_\alpha J_5^\alpha=-\frac{n_f\alpha_s}{4\pi} F^{\mu\nu}\tilde{F}_{\mu\nu}.
\eeq
This leads to the following  exact relation:
\beq
2Mg_A(t) + \frac{tg_P(t) }{2M}
=i\frac{\langle P_2|\frac{n_f\alpha_s}{4\pi} F\tilde{F}|P_1\rangle}{\bar{u}(P_2)\gamma_5u(P_1)} \, .\label{exact2}
\eeq 
We see that, in the absence of the  anomaly (i.e., if the right-hand side were zero), $g_P(t)$ would have a  pole  at $t=0$,
\beq
\frac{g_P(t)}{2M} \approx  -\frac{2M \Delta \Sigma}{t} \qquad (t\to 0) ,\label{gp}
\eeq
where $\Delta \Sigma=g_A(0)$ is the quark helicity contribution to the nucleon spin. Such a pole can be interpreted as due  to the exchange of the massless ninth Nambu-Goldstone boson, the `primordial' $\eta_0$ meson. Moreover,  (\ref{gpint}) suggests that already the flavor-singlet GPD $\sum_q \tilde{E}_q$ would  have a pole $1/t$, just like (\ref{pipole}).  

In reality, however, the U$_A$(1) axial symmetry is explicitly broken due to the anomaly, and $g_P(t)$ exhibits a pole at the physical $\eta'$ meson mass $t=m^2_{\eta'}$. The exact mechanism behind this scenario was a great  debate  in the late 70s culminating in the works of Witten \cite{Witten:1979vv} and Veneziano \cite{Veneziano:1979ec}. In a nutshell, $\eta_0$ acquires mass via a resummation \cite{Veneziano:1979ec}
\beq
\frac{1}{t}+ \frac{m_{\eta'}^2}{t^2} + \frac{m_{\eta'}^4}{t^3}+\cdots =\frac{1}{t-m_{\eta'}^2}  = -\left(   \frac{1}{t}\frac{m_{\eta'}^2}{m_{\eta'}^2-t}-\frac{1}{t}\right) , \label{resum}
 \eeq
due to its coupling with the gluonic topological fluctuations  $m_{\eta'}^2 \propto \langle (F\tilde{F})^2\rangle$. On the right-hand side, we have deliberately expressed the resulting  propagator as the difference of two poles at $t=0$. Now let us compare this with (\ref{exact2}) which can be identically rewritten in the form 
\beq
\frac{g_P(t)}{2M}&=& \frac{1}{t}\left(i\frac{\langle P_2|\frac{n_f\alpha_s}{4\pi} F\tilde{F}|P_1\rangle}{\bar{u}(P_2)\gamma_5u(P_1)}-2Mg_A(t) \right) \nn 
&= & \frac{1}{t} \left( i\frac{\langle P_2|\frac{n_f\alpha_s}{4\pi} F\tilde{F}|P_1\rangle}{\bar{u}(P_2)\gamma_5u(P_1)}-i\left.\frac{\langle P_2|\frac{n_f\alpha_s}{4\pi} F\tilde{F}|P_1\rangle}{\bar{u}(P_2)\gamma_5u(P_1)}\right|_{t=0} \right)+2M\frac{g_A(0)-g_A(t)}{t}\, .
\label{cf2}
\eeq
We neglect the last term assuming $g_A(t)\approx g_A(0)=\Delta\Sigma$ to be varying only slowly with $t$.\footnote{A partial justification of this comes from the large-$N_c$ approximation where $m_{\eta'}\sim {\cal O}(1/\sqrt{N_c})$ is considered as small, at least parametrically, compared to the singlet axial vector meson masses  $m_A\sim {\cal O}(N_c^0)$. Thus, as long as one is interested in the region $|t|\sim m_{\eta'}^2$, the variation of $g_A(t)\sim 1/(t-m_A^2)$ can be neglected.  In  practice, however, the $\eta'(957)$ is only slightly lighter than the $f_1(1285)$. } The right-hand side can then be interpreted as a cancellation of two poles at $t=0$, just like (\ref{resum}), between  the `anomaly pole' (first term) and  the naive pole (\ref{gp}) from the massless $\eta_0$ meson exchange (second term). 
Eqs. (\ref{resum}) and (\ref{cf2}) are actually identical in the single-pole approximation  where (\ref{cf2}) is saturated by 
\beq
\frac{g_P(t)}{2M} \approx  \frac{2M\Delta\Sigma}{m_{\eta'}^2-t} \qquad i\frac{\langle P_2|\frac{n_f\alpha_s}{4\pi} F\tilde{F}|P_1\rangle}{\bar{u}(P_2)\gamma_5u(P_1)} \approx 2M\Delta\Sigma \frac{m_{\eta'}^2}{m_{\eta'}^2-t} . \label{singlepole}
\eeq

In the context of polarized DIS, the cancellation of poles just described had been originally envisaged in  \cite{Jaffe:1989xy} and further elaborated in  \cite{Tarasov:2021yll} to resolve issues with the $g_1$ structure function. Compton scattering and GPDs offer a  more general setup to explore the physics of the anomaly to its full extent.

\subsubsection{Gravitational form factors}

We now point out that  one can repeat the same story for the QCD energy momentum tensor $\Theta^{\alpha\beta}$ and its nucleon matrix element that defines 
the gravitational form factors,
\beq
\langle P_2|\Theta^{\alpha\beta}|P_1\rangle &=& \bar{u}(P_2)\left[A(t)\frac{P^{\alpha}P^{\beta}}{M}+(A(t)+B(t))\frac{P^{(\alpha}i\sigma^{\beta)\lambda}l_\lambda}{2M} + D(t)\frac{l^\alpha l^\beta-g^{\alpha\beta}t}{4M}\right]u(P_1) ,\label{gff}
\eeq
where $a^{(\mu}b^{\nu)} =\frac{1}{2}(a^\mu b^\nu + a^\nu b^\mu)$. 
Taking the trace, we find an exact constraint among the form factors:
\beq
\langle P_2|(\Theta)^{\alpha}_{\alpha}|P_1\rangle = 
 M\left(A(t)+\frac{B(t)}{4M^2}t  - \frac{3D(t)}{4M^2}t\right)\bar{u}(P_2)u(P_1) = \langle P_2|\frac{\beta(g)}{2g}F^{\mu\nu}F_{\mu\nu}|P_1\rangle. \label{trace}
\eeq
The right-hand side, on which 
$\beta(g)$ is the QCD beta function, is the trace anomaly which signifies the explicit breaking of conformal symmetry. If one naively neglects it, one finds a massless pole in $D(t)$ at $t=0$:
\beq
\frac{3D(t) }{4}\approx \frac{M^2}{t} \qquad (t\to 0), \label{dpo}
\eeq
where the conditions $A(0)=1$ and $B(0)=0$ have been used (so that one can omit $tB(t)$ as $t\to 0$). By analogy with  the massless $\eta_0$ pole in (\ref{gp}), one might  interpret the pole in (\ref{dpo}) as due to the exchange of spin-0 glueballs which  would couple to the operator $\Theta^{\alpha\beta}$ and which would have been massless in the absence of the trace anomaly \cite{Bhattacharya:2022xxw}. In reality, however, the anomaly modifies   (\ref{dpo}) to  
\beq
\frac{3D(t)}{4}  &\approx& \frac{M^2}{t}\left(A(t)-\frac{\langle P_2|\frac{\beta(g)}{2g}F^2|P_1\rangle}{M\bar{u}(P_2)u(P_1)} \right)\nn 
&=& -\frac{M}{t} \left(  \frac{\langle P_2|\frac{\beta(g)}{2g}F^2|P_1\rangle}{\bar{u}(P_2)u(P_1)} -\left.\frac{\langle P_2|\frac{\beta(g)}{2g}F^2|P_1\rangle}{\bar{u}(P_2)u(P_1)}\right|_{t=0}\right)+M^2\frac{A(t)-A(0)}{t}. \label{parti}
\eeq
Note the  similarity to  (\ref{cf2}). The $D(t)$-form factor can be interpreted as the difference of two poles at $t=0$, between the `anomaly pole' (first term in the brackets) and the naive glueball pole  (\ref{dpo}) (second term in the brackets). As a result of this cancellation, the pole in $D(t)$ is shifted from $t=0$ to physical glueball masses $t=m_G^2$ presumably in a way similar to (\ref{resum}). 
However, unlike the situation in (\ref{cf2}), in the present case   the last term $A(t)-A(0)$ of (\ref{parti}), which is related to spin-2 glueballs \cite{Bhattacharya:2022xxw}, is likely important at least from the
large-$N_c$ perspective. Since the trace anomaly cannot be turned off in the large-$N_c$ limit, the masses of $2^{++}$ and $0^{++}$ glueballs are both  ${\cal O}(N_c^0)$. Moreover, the analysis in  \cite{Fujita:2022jus} suggests that the single pole approximation (cf., (\ref{singlepole})) may not be  a good approximation. The $D(t)$-form factor thus exhibits  `glueball dominance' 
\beq
D(t) = \sum_i^{0^{++}} \frac{a_i}{m_{G_i}^2-t} + \sum_j^{2^{++}} \frac{b_j}{m_{G_j}^2-t},
\eeq
where the two contributions come from the $\langle F^2\rangle$ and $A(t)$ terms in (\ref{parti}), respectively.  
Incidentally, by taking the $t\to 0$ limit of (\ref{parti}), one finds \cite{Cebulla:2007ei}
\beq
 \frac{3D(0)}{4} = - M\left.\frac{d}{dt} \frac{\langle P_2|\frac{\beta(g)}{2g}F^2|P_1\rangle}{\bar{u}(P_2)u(P_1)}\right|_{t=0}+\left.M^2\frac{dA(t)}{dt}\right|_{t=0}. \label{slope}
\eeq
The slope of a form factor at $t=0$ defines a `radius' of the hadron. Eq.~(\ref{slope}) shows that the D-term $D(t=0)$ is related to the difference between two radii, one defined by  the scalar form factor $\langle F^2\rangle$ (related to the $0^{++}$ glueball masses) and the other by the $A$-form factor (related to the $2^{++}$ glueball masses), see recent discussions in  \cite{Ji:2021mtz,Mamo:2021krl,Kharzeev:2021qkd,Fujita:2022jus}. 

As we have seen in the above three examples, the existence or not of a massless pole in form factors teaches us fundamental insights into the nonperturbative dynamics of QCD. However, despite the known connections between form factors and GPDs, the corresponding discussion at the GPD level has been limited to the isovector sector (\ref{pipole}) in the literature (see, however, \cite{Bass:2001dg}). Our main purpose is to extend this argument to the singlet sector.

\subsection{Anomaly poles in GPDs}

Let us now return to our context.  We have argued in Section V that the pole $1/t$ in the one-loop Compton amplitude (\ref{asym}) should be absorbed into $\tilde{E}_q$. This means that $\tilde{E}_q$ acquires a component related to the twist-four GPD $\tilde{\cal F}$:
\beq
\sum_q (\tilde{E}_q(x,\xi,t)+\tilde{E}_{q}(-x,\xi,t)) = \frac{T_R n_f\alpha_s }{\pi} \frac{M^2}{t} \tilde{C}^{\rm anom}\otimes \tilde{\cal F}(x,\xi,t) +\cdots, \label{relation}
\eeq 
where $\tilde{C}^{\rm anom}$ is defined in (\ref{white2}). 
Integrating over $x$, we exactly reproduce the first term of (\ref{cf2}). Moreover, (\ref{cf2})  suggests that there is another,  `primordial' pole in $\tilde{E}_q$ which exactly cancels the $1/t$ pole to make $\tilde{E}_q$ finite for all values of $x$ and $\xi$ in the limit $t\to 0$. A simple, yet ad-hoc fix consistent with (\ref{cf2}) is to add a `counterterm' 
\beq
\sum_q (\tilde{E}_q(x,\xi,t)+\tilde{E}_{q}(-x,\xi,t)) \approx \frac{T_R n_f\alpha_s }{\pi} \frac{M^2}{t} \tilde{C}^{\rm anom}\otimes (\tilde{\cal F}(x,\xi,t)-\tilde{\cal F}(x,\xi, 0)). \label{2-4}
\eeq
This may be viewed as the non-local version of the local relation (\ref{cf2}). The second, added term is an analog of (\ref{pipole}), but interestingly, in the present case the pole is not limited to the ERBL region $x< \xi$. 
We postulate (\ref{2-4}) as  a nonperturbative relation between the twist-two and twist-four GPDs mediated by the chiral anomaly.

The fate of the $1/t$ pole in the unpolarized sector and its connection to the trace anomaly are  more involved. This is partly  because the QCD energy momentum tensor consists of a quark and a gluon part, $\Theta^{\alpha\beta}=\sum_q \Theta_q^{\alpha\beta}+\Theta_g^{\alpha\beta}$, in contrast to $J_5^\alpha$ which is purely a quark operator.  Accordingly, one can define gravitational form factors separately for quarks and gluons \cite{Ji:1996ek}:
\beq
\langle P_2|\Theta_{q,g}^{\alpha\beta}|P_1\rangle &=& \bar{u}(P_2)\left[A_{q,g}(t)\frac{P^{\alpha}P^{\beta}}{M}+(A_{q,g}(t)+B_{q,g}(t))\frac{P^{(\alpha}i\sigma^{\beta)\lambda}l_\lambda}{2M} + D_{q,g}(t)\frac{l^\alpha l^\beta-g^{\alpha\beta}t}{4M} + \bar{C}_{q,g} (t)M g^{\alpha\beta}\right]u(P_1).\label{gff2} \nn
\eeq
They  are related to the second moments of the unpolarized quark GPDs,
\begin{align}
\int_{-1}^1 dx x H_q(x,\xi,t) & =\int_0^1 dx x  (H_q(x,\xi,t)-H_q(-x,\xi,t))   = A_q(t) + \xi^2 D_q(t) \, ,\label{poly0} \\
\int_{-1}^1 dx x E_q(x,\xi,t) & = \int_0^1 dx x (E_q(x,\xi,t)-E_q(-x,\xi,t))  =  B_q(t) - \xi^2 D_q(t) \,,
\end{align}
and similarly for the gluon GPDs. 
Taking the trace of (\ref{gff}), we find 
\beq
\langle P_2|\sum_q(\Theta_q)^{\alpha}_{\alpha}|P_1\rangle &=& 
 \sum_q M\left(A_q(t)+4\bar{C}_q(t)+\frac{B_q(t)}{4M^2}t  - \frac{3D_q(t)}{4M^2}t\right)\bar{u}(P_2)u(P_1) \nn
 &=&\langle P_2|\frac{\beta_q(g)}{2g}F^2|P_1\rangle \approx \langle P_2|\frac{T_Rn_f \alpha_s}{6\pi}F^2|P_1\rangle, \label{trace2}
\eeq
where $\frac{\beta_q}{2g}$ is the quark part of the trace anomaly that can be systematically calculated in perturbation theory  \cite{Hatta:2018sqd,Tanaka:2018nae,Ahmed:2022adh,Tanaka:2022wzy}.   To lowest order, it is simply the $n_f$ term of the beta function:
\beq
 (\Theta_q)^\alpha_\alpha +(\Theta_g)^\alpha_\alpha = \frac{\beta(g)}{2g}F^{\mu\nu}F_{\mu\nu} =-\frac{\alpha_s}{8\pi}\left( \frac{11N_c}{3}-\frac{4T_Rn_f}{3} \right)  F^2+\cdots. \label{tra}
\eeq 
Clearly, (\ref{trace2}) is not as constraining as (\ref{trace}) because of the new form factors $B_q(t), \bar{C}_q(t)$. (Note that $B_q(0),\bar{C}_q(0)\neq 0$ although  $B_q(0)+B_g(0)=\bar{C}_q(t)+\bar{C}_g(t)=0$.) Nevertheless we may try to rewrite it in a way similar to (\ref{parti})
\beq
\sum_q \frac{3D_q(t)-B_q(t)}{4} =  -\frac{M}{t}\left( \frac{\langle P_2|\frac{\beta_q}{2g}F^2|P_1\rangle}{\bar{u}(P_2)u(P_1)} -\left.\frac{\langle P_2|\frac{\beta_q}{2g}F^2|P_1\rangle}{\bar{u}(P_2)u(P_1)}\right|_{t=0}\right)+\frac{M^2}{t}\sum_q \Bigl(A_q(t)+4\bar{C}_q(t)-A_q(0)-4\bar{C}_q(0)\Bigr). \label{trace3}
\eeq

Let us  now discuss how the constraint  (\ref{trace3}) from the trace anomaly is encoded in the   GPDs. We have argued that the anomaly poles in (\ref{sym}) should be absorbed into the unpolarized GPDs,
\begin{align}
\sum_q (H_q(x,\xi,t)-H_q(-x,\xi,t)) & =  \frac{T_Rn_f\alpha_s}{\pi}\frac{M^2}{t} C^{\rm anom}\otimes  {\cal F}(x,\xi,t)+\cdots, \\
\sum_q (E_q(x,\xi,t)-E_q(-x,\xi,t)) & = - \frac{T_Rn_f\alpha_s}{\pi}\frac{M^2}{t} C^{\rm anom}\otimes {\cal F}(x,\xi,t) 
+\cdots, \label{absorb}
\end{align}
where $C^{\rm anom}$ is defined in (\ref{white}).  
Taking the second moment and comparing with (\ref{poly0}), we find
\beq
\left.\sum_q A_q(t)\right|_{\rm pole} =-\left.\sum_q B_q(t)\right|_{\rm pole} =   -\frac{M}{t} \frac{\langle P_2| \frac{T_Rn_f \alpha_s}{6\pi }F^{\mu\nu}(i\overleftrightarrow{D}^+)^2F_{\mu\nu}|P_1\rangle}{(P^+)^2\bar{u}(P_2)u(P_1)},  
\label{apole}
\eeq
\beq
  \left.\sum_q D_q(t) \right|_{\rm pole}= -\frac{M}{t} \frac{\langle P_2|\frac{T_Rn_f \alpha_s}{6\pi } F^2|P_1\rangle}{\bar{u}(P_2)u(P_1)} .
  \label{dpole}
\eeq
Eq.~(\ref{dpole}) seems to reproduce the first term of (\ref{trace3}) after $\beta_q$ is expanded to lowest order. However, apparently there is a factor $\frac{3}{4}$ mismatch in the normalization. Besides, the previous argument around (\ref{parti}) did not hint at the possible existence of an anomaly  pole in the $A_q,B_q$ form factors. 

In order to understand these differences, we quote the one-loop result for the energy momentum tensor matrix element  
between on-shell gluon (not nucleon) states
\cite{Giannotti:2008cv}:
\beq
\langle p_2|\Theta_q^{\alpha\beta}|p_1\rangle = -\frac{T_R\alpha_s}{6\pi } \left(\frac{p^\alpha p^\beta}{t} + \frac{l^\alpha l^\beta-t g^{\alpha\beta}}{4t}\right)\langle p_2|F^{\mu\nu}F_{\mu\nu}|p_1\rangle +\cdots \, . \label{qed}
\eeq
 A superficial comparison with (\ref{gff2}) suggests that poles of equal magnitude are induced in the $A_q,B_q,D_q$  form factors  
\beq
A_q(t)\approx  -B_q(t)\approx D_q(t)\sim \frac{\langle \alpha_s F^2\rangle}{t},
\eeq
and the issue of the factor $\frac{3}{4}$ goes away because $\frac{3D_q(t)-B_q(t)}{4}\approx D_q(t)$ on the left hand side of (\ref{trace3}). 
Taking the trace of (\ref{qed}), we find 
\beq
\langle p_2|(\Theta_q)^\alpha_\alpha|p_1\rangle = \langle p_2|\frac{T_R\alpha_s}{6\pi}F^2|p_1\rangle, \label{pert}
\eeq
which is the correct trace anomaly relation to this order. To obtain (\ref{pert}), it is important to use the on-shell condition $p^2=-t/4$ of the external states, so that the two terms in (\ref{qed}) contribute $\frac{1}{4}$ and $\frac{3}{4}$ of the total anomaly, respectively. Going from gluon to nucleon targets, we see that the way the trace anomaly relation (\ref{trace2}) is fulfilled among various form factors is highly nontrivial. A different, spin-2 operator $F(D^+)^2F$ is involved in the $A_q,B_q$ form factor (\ref{apole}) due to the convolution integral in $x$. Moreover, a naive identification $p^\alpha p^\beta \to P^\alpha P^\beta$ is precarious because the nucleon is massive $P^2=M^2-t/4$. While the difference is negligible when $\sqrt{|t|}\gg M$, this obscures the fate of the poles in (\ref{apole}) as $t$ gets smaller.

On the other hand, the tensor $l^\alpha l^\beta$ is formally identical for both the nucleon and gluon targets,  $l^\alpha = p_2^\alpha-p_1^\alpha=P_2^\alpha -P_1^\alpha$. We may therefore expect that the anomaly relation at the partonic level is better reflected in the $D_q(t)$ form factor even at the hadronic level, just like the $g_P(t)$ form factor which is the coefficient of $l^\alpha$. Indeed, the opposite signs in (\ref{absorb}) suggests that the pole terms  mainly feed into the so-called Polyakov-Weiss D-term \cite{Polyakov:1999gs} of the unpolarized GPDs,
\beq
H^{\rm PW}_q(x,\xi,t)= -E^{\rm PW}_q(x,\xi,t) = \theta(\xi-|x|)D_q(x/\xi,t).
\eeq
The distribution $D_q(z,t)$ is odd in $z$ and is solely responsible for the highest power of $\xi$ in the moments of  GPDs.  
In order to extract it, we take the $n$-th moment of  (\ref{absorb}) with odd integers $n$, 
\beq
\sum_q\int_{-1}^1 dx x^n H_q(x,\xi,t) 
&\approx &\frac{T_Rn_f\alpha_s}{\pi}\frac{M^2}{t}\int_0^1 dx \frac{x^n}{(n+2)(n+3)}\frac{1-\left(\frac{\xi}{x}\right)^{n+3}}{1-\frac{\xi^2}{x^2}} \left({\cal F}(x,\xi,t)-{\cal F}(x,\xi,0)\right) \nn 
&\equiv & \sum_{i=0}^{n+1} \sum_q h_{qn}^i \xi^i, \label{weis}
\eeq
 where we have minimally subtracted  the pole at $t=0$ as in (\ref{2-4}).  The highest power $h_{qn}^{n+1}$ is related to  $D_q(z,t)$ as   
\beq
\sum_q \int_{-1}^1dz z^n D_q(z,t) &\approx & \sum_q h^{n+1}_{qn}(t) \nn 
&=&  \frac{T_Rn_f\alpha_s}{\pi}\frac{M^2}{t}\frac{1}{(n+2)(n+3)}\int_0^1 \frac{dx}{x}  \left({\cal F}(x,\xi,t)-{\cal F}(x,\xi,0)\right)\nn 
 &=& -2 \frac{T_R n_f\alpha_s}{\pi}\frac{M}{t} \frac{1}{(n+2)(n+3)} \left( \frac{\langle P_2|F^2|P_1\rangle}{\bar{u}(P_2)u(P_1)} -\left.\frac{\langle P_2|F^2|P_1\rangle}{\bar{u}(P_2)u(P_1)}\right|_{t=0}\right).
 \label{mel}
\eeq
By definition,  the $n=1$ moment is the gravitational form factor $\int_{-1}^1dz z D_q(z,t) = D_q(t)$. 
Inverting the Mellin transform (\ref{mel}) and noting that $D_q(z,t)$ is an odd function of $z$, we obtain
\beq
\sum_q D_q(z,t) \approx -\frac{T_Rn_f \alpha_s }{\pi}z(1-|z|)\frac{M}{t}\left(\frac{\langle P_2|F^2|P_1\rangle}{\bar{u}(P_2)u(P_1)} -\left.\frac{\langle P_2|F^2|P_1\rangle}{\bar{u}(P_2)u(P_1)}\right|_{t=0} \right), \label{pw}
\eeq
and in particular,
\beq
\sum_q D_q(t) \approx -\frac{M}{t}\left(\frac{\langle P_2|\frac{T_Rn_f \alpha_s }{6\pi}F^2|P_1\rangle}{\bar{u}(P_2)u(P_1)} -\left.\frac{\langle P_2|\frac{T_Rn_f \alpha_s }{6\pi}F^2|P_1\rangle}{\bar{u}(P_2)u(P_1)}\right|_{t=0} \right). \label{pw2}
\eeq
Since $\langle P|F^2|P\rangle<0$ in QCD and the form factor $\langle P_2|F^2|P_1\rangle$ is a decreasing function of $|t|$, the right-hand side of (\ref{pw}) is positive, whereas $D_q(t)$ is usually believed to be negative. While we expect that eventually the leading-order coefficient $\frac{T_Rn_f\alpha_s}{6\pi}$ will be replaced by  $\frac{\beta_q}{2g}$ after including higher-order corrections, according to the three-loop analyses in \cite{Tanaka:2018nae,Metz:2020vxd,Ahmed:2022adh,Tanaka:2022wzy},  the sign does not flip $\frac{\beta_q}{2g}>0$. This suggests that the other terms in (\ref{trace3}) that were neglected in the above minimal subtraction procedure may be numerically important as we already suspected in the argument below (\ref{parti}). Note also that the sign does flip if one includes the gluon contribution to recover the full beta function of QCD $\frac{\beta_q}{2g}\to \frac{\beta}{2g}<0$.

\section{Conclusions}

In this work, we have performed a complete one-loop calculation of the Compton scattering amplitude using momentum transfer $t$ as the regulator of the collinear singularity.  Our approach differs from all the previous calculations in the GPD literature where one typically uses dimensional regularization to isolate the collinear singularity and sets $t=0$ right from the start, assuming that nonzero $t$ only generates higher-twist corrections of order $t/Q^2$. In practice, the introduction of an additional variable $t$ makes the calculation more cumbersome and brings in unusual features. In the gluon initiated channel, we have found anomaly poles $1/t$  (\ref{a1}), (\ref{a2}) accompanied by twist-four GPDs (\ref{four}), (\ref{two}) in both the real and imaginary parts of the Compton amplitude, confirming and extending our previous finding \cite{Bhattacharya:2022xxw}. In the quark initiated channel, we have unexpectedly found uncancelled single and double IR poles in the `coefficient functions' (\ref{cq}), (\ref{cq2}). Each of these features potentially implies the violation of factorization. 
However, we have also performed the one-loop calculation of GPDs for quark and gluon states with the same set of regulators and showed how all these poles can be systematically absorbed into the GPDs themselves. This shows that QCD factorization is restored at least to this order. 

This is however not the end of the story. 
We have also explored connections between GPDs and  anomalies, as a natural and necessary consequence of the known connections between form factors and anomalies. We have argued that once the poles $1/t$ have been absorbed into  GPDs, they become a part of the GPDs. In other words, anomalies nonperturbatively relate twist-two and twist-four GPDs. Such  relations, once integrated over $x$, are expected to reproduce the constraints among the corresponding form factors. This scenario seems to be working  for the polarized GPD $\tilde{E}_q$ and its connection to the chiral anomaly. Relation (\ref{2-4}), partly supported by the large-$N_c$ argument, can be viewed as the $x$-dependent generalization of the form factor relation (\ref{exact2}). The situation is more complicated (and more interesting) for the unpolarized GPDs $H_q,E_q$ and their relation to the trace anomaly.  We have argued that the anomaly mostly constrains the $D_q(t)$ form factor and its GPD analog, the Polyakov-Weiss D-term. The results we have arrived at (\ref{pw}) (\ref{pw2}) are roughly consistent with the anomaly relation (\ref{trace3}), but they differ in detail. Further investigation in this direction is necessary.  

In conclusion, we have proposed  finite-$t$ regularization as an alternative factorization scheme that elucidates the physics of anomalies. This is a scheme where we are able to unravel novel connections between twist-two and twist-four GPDs mediated by the  anomalies of QCD. Admittedly, the calculation is more cumbersome than the standard dimensional regularization with $t=0$. Still, the chiral and trace anomalies  are among the most fascinating  phenomena of QCD with far-reaching consequences,  and we believe that research on GPDs is enriched by incorporating such fundamental problems.  
There are a number of directions along which the current work can be refined or extended, in addition to the aforementioned tension between  (\ref{trace3}) and (\ref{pw2}). First, we strongly suspect that anomaly poles are present in higher order perturbation theory. Especially in the symmetric case, we expect that each additional loop provides the corresponding term in the expansion of the (quark part of the)  beta function. A related question is whether there are  anomaly poles in the {\it gluon} GPDs that complement the quark ones to restore the full  beta function  $\beta=\beta_q+\beta_g$ \cite{Hatta:2018sqd}. Another important question which has not been addressed at all in this paper is how to understand the new relations from a renormalization group point of view. In the present scheme, the mixing between the twist-two and twist-four GPDs occurs as a result of a finite subtraction rather than the DGLAP evolution of GPDs.  The UV properties of the twist-four GPDs  (\ref{four}) and (\ref{two}) have been studied in \cite{Hatta:2020iin,Hatta:2020ltd,Radyushkin:2021fel,Radyushkin:2022qvt}, but more work is certainly needed. Furthermore, it is well known that at twist-3 accuracy, the amplitude for DVCS off the nucleon contains twist-3 GPDs apart from the usual twist-2 GPDs. It is also interesting to  pursue whether or not there are imprints of anomalies on twist-3 GPDs and related observables. 
 Finally, constraints  from anomalies should be implemented in  the modeling of GPDs. In particular, the specific functional form given in (\ref{pw}) might be helpful to model this poorly constrained distribution.

\begin{acknowledgements}
We are very grateful to Vladimir Braun and Anatoly Radyushkin for many useful discussions. We also thank Kornelija Passek-Kumeri$\check{\rm c}$ki, Swagato Mukherjee, Kazuhiro Tanaka, Raju Venugopalan and Christian Weiss for   discussions. S.~B. and Y.~H. are supported by the U.S. Department of Energy under Contract No. DE-SC0012704, and   Laboratory Directed Research and Development (LDRD) funds from Brookhaven Science Associates. Y.~H. is also supported by the framework of the Saturated Glue (SURGE)
Topical Theory Collaboration. W.~V. has been supported by Deutsche Forschungsgemeinschaft (DFG) through the Research Unit FOR 2926 (project 409651613). 
\end{acknowledgements}

\appendix

\section{Derivation of Eq.~(\ref{dg})}

In this appendix we give an outline of the derivation of (\ref{dg}) and (\ref{finunpol}). The other results in Section \ref{gpd1loop} can be derived similarly. For the quark matrix elements, we work in the Feynman gauge. 
The ladder diagram reads, up to a prefactor, 
\beq
\mu^{2\epsilon}\int \frac{dk^- d^{2-2\epsilon}k_\perp}{(2\pi)^{3-2\epsilon}} \bar{u}(p+l/2)\frac{\gamma_\mu (\Slash k+\Slash l/2)\gamma^+(\Slash k-\Slash l/2)\gamma^\mu}{(p-k)^2(k-l/2)^2(k+l/2)^2} u(p-l/2), \label{lad}
\eeq
where 
\beq
k^+=xp^+, \quad l^+=-2\xi p^+, \quad l^-=-\frac{\xi l^2}{4p^+}, \quad \vec{l}_\perp^2=(\xi^2-1)l^2, \quad p^-=-\frac{l^2}{8p^+}.
\eeq
In the DGLAP region $\xi<x<1$, the $k^-$ integral can be done by picking up the pole of $(p-k)^2+i\epsilon=0$
 at 
\beq
k^- =-\frac{k_\perp^2+\frac{1-x}{4}l^2}{2(1-x)p^+} ,\label{use}
\eeq
in the upper half plane. The remaining propagators can be combined as 
\beq
\frac{1}{1-x}\int_0^1da \frac{A(k_\perp)}{(k^2+(1-2a)k\cdot l+\frac{l^2}{4})^2} 
=\frac{1-x}{(1-\xi(1-2a))^2}\int_0^1 da \frac{A(k'_\perp -\frac{(1-2a)(1-x)}{2(1-\xi(1-2a))}l_\perp)}{\left(k'^2_\perp-\frac{(1-a)a(1-x)^2l^2}{(1-\xi(1-2a))^2}\right)^2}, 
\eeq
where in the denominator we have shifted momentum $k_\perp \to k'_\perp$ to complete the square. In the numerator we have projected onto the twist-two component,
\beq
\bar{u}(p+l/2)\cdots u(p-l/2)\to \left[(1-\epsilon)k'^2_\perp +(B-\epsilon C)l^2\right]\bar{u}\gamma^+u\equiv  A\bar{u}\gamma^+u, \label{a(k)}
\eeq
with
\beq
B=\frac{(1-\xi + a(x+2\xi-1))(x-\xi-a(x-2\xi-1))}{(1-\xi(1-2a))^2},\quad  C=\frac{(1-a)a(1-x)^2}{(1-\xi(1-2a))^2}.
\eeq
The terms linear in $k'_\perp$ have been dropped in (\ref{a(k)}) since they vanish after the $k'_\perp$ integral:
\beq
&&\mu^{2\epsilon}\int_0^1 da \frac{1-x}{(1-\xi(1-2a))^2} \int \frac{d^{2-2\epsilon}k'_\perp}{(2\pi)^{2-2\epsilon}}  \frac{ (1-\epsilon)k'^2_\perp+(B-\epsilon C)l^2}{\left(k'^2_\perp-Cl^2\right)^2} \nn
&& \approx \left(\frac{ \mu^2}{-l^2}\right)^{\epsilon}\int_0^1 da \frac{1-x}{(1-\xi(1-2a))^2} \frac{1}{(4\pi)^{1-\epsilon}}\left(  \frac{ (1-2\epsilon)\Gamma(\epsilon)}{C^\epsilon}-\frac{B\Gamma(1+\epsilon)}{C^{1+\epsilon}} \right).
\eeq
The first integral gives a UV pole:
\beq
&&\frac{\Gamma(\epsilon_{\UV})}{4\pi}\left(\frac{4\pi \mu^2}{-l^2}\right)^{\epsilon} \int_0^1 da \frac{(1-x)(1-2\epsilon)}{(1-\xi(1-2a))^2C^\epsilon} \nn
&&=\frac{1}{4\pi }\left(\frac{ \tilde{\mu}^2}{-l^2}\right)^{\epsilon}\left(\frac{1}{\epsilon_{\UV}}\frac{1-x}{1-\xi^2} - \frac{(1-x)(2\ln (1-x)-\ln (1-\xi^2))}{1-\xi^2}\right),\label{lead}
\eeq
while 
the second integral gives double and single IR poles:
\beq
&&\left(\frac{4\pi\mu^2}{-l^2}\right)^{\epsilon}\int_0^1 da \frac{1-x}{(1-\xi(1-2a))^2}\frac{-B\Gamma(1+\epsilon)}{C^{1+\epsilon}} \nn
&=& \left(\frac{\tilde{\mu}^2}{-l^2}\right)^{\epsilon} \frac{1}{(1-x)^{1+2\epsilon}}\left(\frac{2(x-\xi^2)}{1-\xi^2}\frac{1}{\epsilon_{\IR}} -\frac{(1-x)^2-2(x-\xi^2)\ln (1-\xi^2)}{1-\xi^2} + \epsilon f(x) \right) \nn
&=&  \left(\frac{\tilde{\mu}^2}{-l^2}\right)^{\epsilon} \Biggl[-\delta(1-x) \left(\frac{1}{\epsilon_{\IR}^2} + \frac{\ln (1-\xi^2)}{\epsilon_{\IR}}\right) + \frac{2(x-\xi^2)}{(1-x)_+(1-\xi^2)}\frac{1}{\epsilon_{\rm IR}}- \frac{4(x-\xi^2)}{1-\xi^2}\left(\frac{\ln(1-x)}{1-x}\right)_+ \nn
&& - \frac{(1-x)^2-2(x-\xi^2)\ln (1-\xi^2)}{(1-\xi^2)(1-x)_+} -\frac{1}{2}\delta(1-x)\left( \ln^2(1-\xi^2) -\frac{\pi^2}{6}\right)  \Biggr]. \label{14}
\eeq
Here $f(x)$ is a certain function whose value at $x=1$ is the only thing we need.

In Feynman gauge, there are two other diagrams, giving
\beq
\int \frac{d^dk}{(2\pi)^{d-1}} \bar{u}(p+l/2) \frac{\gamma^+\Slash k \gamma^+  \bigl(\delta(k^+-(x+\xi)p^+)-\delta((1-x)p^+)\bigr) }{k^2(p-l/2-k)^2((1+\xi)p^+ -k^+)}u(p-l/2), \label{delta}
\eeq
and the corresponding contribution for its mirror diagram. They can be similarly evaluated. The result is 
\beq
\left(\frac{2(x-\xi^2)}{(1-\xi^2)(1-x)_+} + \delta(1-x)\left(2 -\ln (1-\xi^2)\right)\right)\frac{\left(\frac{\tilde{\mu}^2}{-l^2}\right)^\epsilon}{4\pi} \left(\frac{1}{\epsilon_{\UV}}-\frac{1}{\epsilon_{\IR}}\right) .
\eeq
Adding also the quark self energy diagrams on the external legs, we arrive at (\ref{dg}) and (\ref{finunpol}). 
We note that the calculation can also be performed in Landau gauge which has the advantage that the self-energy
diagrams vanish identically. 

In the ERBL region $\xi>x$, the pole of $(k+l/2)^2$ in (\ref{lad}) moves to the upper half plane. We thus pick up the pole of $(k-l/2)^2$ at 
\beq
k^-_c=\frac{k_\perp^2-(x\xi+1)\frac{l^2}{4}-\vec{k}_\perp\cdot \vec{l}_\perp}{2(x+\xi)p^+}, \label{er}
\eeq
in the lower half plane. The first term in (\ref{delta}) also contributes (but not its mirror diagram).

\bibliography{references}

\end{document}